\documentclass[preprint,12pt]{elsarticle}
\usepackage{graphicx,psfrag,color}
\usepackage{mathbbol,amsmath,amsfonts,amssymb,bm,dsfont}
\usepackage{wasysym}
\usepackage{amsthm}

\def\m{{\bm{m}}}

\def\dual{\ \stackrel{\Phi_\d}{\longrightarrow}\ }

\def\d{{{\sf d}}}
\def\r{{\bm{r}}}
\def\x{{\bm{x}}}
\def\y{{\bm{y}}}
\def\i{{\bm{e_1}}}
\def\j{{\bm{e_2}}}
\def\k{{\bm{e_3}}}
\def\a{{\bm{a_1}}}
\def\aa{{\bm{a_2}}}
\def\aaa{{\bm{a_3}}}

\def\tr{{\sf Tr}}

\journal{Nuclear Physics B}

\makeindex
\begin{document}
\begin{frontmatter}

\title{Effective and exact holographies \\ from symmetries and dualities}

\author[]{Zohar Nussinov$^1$, Gerardo Ortiz$^{2}$, and Emilio Cobanera$^2$}
\address{$^1$Department of Physics, Washington University, St.
Louis, MO 63160, USA, \\ 
$^2$Department of Physics, Indiana University, Bloomington,
IN 47405, USA.}

\begin{abstract}
The theoretical basis of the phenomenon of effective and exact dimensional reduction, or  holographic
correspondence, is investigated in a wide variety of physical systems. 
We first  derive general inequalities  linking quantum systems of
different spatial  (or spatio-temporal) dimensionality, thus
establishing bounds  on arbitrary correlation functions. These bounds
enforce an {\em effective} dimensional reduction and become most potent
in the presence of certain symmetries. {\em Exact} dimensional reduction
can stem from a duality  that (i) follows from properties of the local
density of states, and/or (ii) from properties of Hamiltonian-dependent
algebras of interactions.  Dualities of the first type (i) are
illustrated with large-$n$ vector theories whose
local density of states  may remain invariant under  transformations
that change the dimension. We argue that  a broad class of examples of
dimensional reduction may be understood in terms of the functional
dependence of observables on the local density of states. Dualities of
the second type (ii) are obtained via {\em bond algebras},
a recently developed  algebraic tool.  We apply this technique to
systems displaying topological quantum  order, and also discuss the 
implications of dimensional reduction for the storage of quantum information. 
\end{abstract}

\end{frontmatter}

\section{Introduction}
Spatial (or space-time) dimensionality, topology, and symmetry are
central concepts in physics. This article studies a phenomenon, that of 
{\em dimensional reduction} or {\em holographic correspondences}, that
highlights how these concepts may come together to produce particularly
striking signatures.

The dimension of a system is one of its most  basic characteristics and
typically represents an important measure of its complexity. 
Dimensional reduction and  holographic correspondences  refer to the
fact  that the ``apparent", or ``obvious" dimension \(D\) of a system
may not be the dimension $d$, \(d<D\), that best characterizes its 
response to experimental probes and its information content. We speak of
dimensional reduction when the  observable physical properties of a
system  in $D$  dimensions  behave as if it were $d$-dimensional. 
We argue for a holographic correspondence when two systems
of different dimensionality seem capable of storing and processing comparable
amounts of information.  Both notions,  often undistinguished in  current
literature, appear in numerous fields including condensed matter, cold
atom, high-energy, and black-hole physics.  Given the cross-disciplinary
nature of the subject there is a need  to better understand the physical
and mathematical basis of dimensional reduction and  holographic
correspondences. We believe that this understanding should be rooted  in
basic principles than in  properties of specific models.

This article constitutes an attempt to study {\it exact} and  {\em
effective} dimensional reductions, and holographic correspondences
within a unified framework. Our approach relies on the recent theory of
dualities \cite{bondprl}, and a new theorem that constrains arbitrary
correlators  of a quantum theory in \(D\) dimensions.  The obtained
inequalities and bounds afford a practical notion of effective
dimensional reduction because they  are specified by homologous
correlators of an associated, {\it local}, \(d<D\) dimensional theory.
Also they become specially useful in the presence of {\it \({\bf
d}\)-dimensional gauge-like symmetries} (or $\bf d$-GLSs for short) 
\cite{NO,BN}, with ${\bf d}<D$, a geometrically distinguished set of
symmetries that we will discuss further below.  The notion of {\it
exact} dimensional reduction is further linked to  the existence of a
duality, mapping the $D$-dimensional theory to a  $d$-dimensional one.
In such situations inequalities may be replaced by exact equalities.  We also introduce the
concept of {\it dimensional reduction  by the density of states}, a
concept that may find its way  to practical applications in the form of
either exact or effective  versions, depending on the problem.  We
illustrate this technique  with $O(n)$ type vector theories.  Finally,
we will show that exact dimensional reduction  is sometimes  connected
to a {\it holographic-like entropy} that scales as the surface of the 
system,  and use large-\(n\) vector theories to illustrate this point.
Remarkably,  our take on dimensional reduction/holographic
correspondences will allow  us to discuss limitations on the storage and
robustness of quantum information. 

While dimensional reduction/holographic correspondences may seem  to be
exotic phenomena, they can in fact appear in many systems as the natural
consequence of a wealth of physical mechanisms:

(1) {\it Restrictions from conservation laws.}  In some cases, a
conservation law (that can be formalized as a gauge constraint) favors
{\it sliding} dynamics along preferred directions  while forbidding
motion along other directions.   For example, the glide principle
\cite{glide} for elastic solids dictates that  dislocations
move preferentially  along directions that are determined by their
topological charges or Burger's vectors.

(2) {\it Reduced kinetics and interactions along one or more
directions.}  In many condensed matter systems a dimensional reduction
can occur due to the {\em confinement} of electrons/atoms along planes
or lines (e.g., wires), whereby certain directions become irrelevant.   
{\em Reduced tunneling}, for instance, can appear along certain directions, 
as in the case of high-$T_{c}$
cuprates and some other layered systems, due to relatively large
separation between planes that suppresses interplane couplings and
tunneling from one layer to the next.  Similar decoupling can  occur in
magnetic systems with {\em spin exchange interactions}  on geometrically
frustrated lattices  \cite{sebastian}, where the coupling between planes
(or other lower-dimensional  subvolumes) along one or more directions
becomes negligible.  In all of these systems, the  effective physical
interactions and/or kinetic hopping matrix elements  describe a
lower-dimensional system. Many of these systems are locally 
inhomogeneous with electrons constrained in some way, 
but at large distances the symmetry between different directions is
restored. That is, in such cases, converse to the Kaluza-Klein type
compactifications that we list next, {\it dimensional reduction may}
(but need not) {\it emerge at local scales}.  At short distance scales
the system is effectively of a lower dimensionality than it is at large
distance scales. 

(3) {\it Kaluza-Klein compactification.} Dimensional reduction can be
achieved geometrically by compactifying $(D-d)$ spatial dimensions of a
theory through the Kaluza-Klein  \cite{Kaluza_Klein} procedure.  At low
energies or temperatures, excitations along these ``small dimensions''
are suppressed, leading to an effective $d$-dimensional theory.  That
is, {\it dimensional reduction emerges at low energies and large
distances}. This type of dimensional reduction is important  in string
\cite{gsw} and supergravity theories, and it has been suggested recently
\cite{langlands,kw} that may also help to address the Langlands duality
which is an open problem in basic mathematics. 

(4) {\it Gravity-gauge dualities {\rm (AdS-CFT)}}.  A seemingly
different, more subtle, and very powerful form of dimensional reduction 
is offered by the  AdS-CFT (or more general  gravity-gauge) holographic
correspondences (dualities). This  correspondence relates, via a
putative strong-weak coupling duality, different theories: string theory
as a theory of quantum gravity on an Anti de Sitter (AdS) background in
$D$ dimensions, and a conformal field theory (CFT) in $d=D-1$
space-time  dimensions.   Since its initial discovery  \cite{maldacena},
the AdS-CFT  correspondence  \cite{maldacena,GKP,Witten,review,nastase}
has been a source of inspiration for many ideas and applications. Apart
from its original use in string theory, as a gravity-gauge
correspondence \cite{maldacena}, it has led to many other applications
that include hadronic hydrodynamics  \cite{hydro} and fundamental
questions relating to transport in strongly correlated  systems 
\cite{electron}. Similar forms of an AdS-CFT type correspondence were
conjectured for cold atom systems with Schr\"odinger type dynamics
\cite{sch}. 

This incomplete list of known mechanisms underscores the multitude of
disparate phenomena that may lead to dimensional reduction and
holographic-like  correspondences.  This article will describe simple
tools that can help address this diversity in a coherent way. As
mentioned already, we distinguish between two scenarios:

(i) {\it Effective dimensional reduction.}  This possibility has, for
the most part, not been explored before (especially in quantum
theories), and forms a cornerstone of this article.  Items (1), (2), and
(3) above  can be seen as realizations of this type of dimensional
reduction. We will illustrate how inequalities with bounds on arbitrary
correlation functions,  typically strengthened by the presence of $\bf
d$-GLSs, relate to an effective dimensional reduction.  

(ii) {\it Exact dimensional reduction.} We identify this scenario with
the existence of dualities connecting  theories of different 
dimensionality. These dualities may be  exact or asymptotically so in
some limit (such as the large-$n$ limit or the weak coupling limit). Case (4)  
is  often conjectured to be exact yet still largely remains to be
proved. The AdS-CFT correspondence has, however, been convincingly
established in the  large-$n$ limit and several related $SU(n)$ matrix
and other models. 

In general, an effective dimensional reduction is  very different in
nature from an {\it exact}  dimensional reduction, whence the  spectrum
of a $D$-dimensional system attains a form akin to that of a  theory in
${d}<D$ dimensions with local interactions.  In this paper, this {\it
effective} scenario is characterized in terms of a flexible formalism
that can describe a range of situations that go from ``very weak" to
``very strong" dimensional reduction. It is essential then to recognize
ingredients that will favor the strongest forms of  effective
dimensional reduction. Symmetries, and more specifically, the ${\bf
d}$-GLSs that we describe in the next section, are one such ingredient. 
In such cases, those symmetries mandate how the bounding $\bf
d$-dimensional correlation  functions decay with distance. 

The concept of exact dimensional reduction (ii) is more restrictive, and
its connection to ${\bf d}$-GLS is less clearly understood.   If
an {\it exact} equivalence is present between all possible correlation
functions in two different theories  then it will be natural  to conjecture that a
duality (i.e., an exact unitary mapping \cite{bondprl}) connects the
theory that displays lower- ($d$) dimensional behavior, yet described as
high- ($D$) dimensional theory, to a dual theory explicitly
$d$-dimensional. Then the inequalities saturate and become equalities. 
This occurs, for instance, in some special models \cite{NO,bond}, such as
Kitaev's  model \cite{kitaev},  that
may be constructed and solved using these unitary mappings. As we will show, this occurs far more
generally in the large-$n$ limit of rather general $O(n)$ type vector
theories. The crux of this result is that in systems such as 
large-$n$  vector theories, the partition function (or generating
functional) of the system  can be expressed  in terms of an effective
density of states functional. Transformations between systems that
reside in different  (space/space-time) dimensions that preserve this
density of states, while maintaining the locality of the theory,
automatically correspond  to (dimension reducing) duality
transformations.

We expect, though, that the most complex cases of dimensional
reduction/holographic correspondences may be understood as the combined
use of (i) and (ii).  

\section{Systems of study}
\label{systems}

The results reported in this work hold for a very broad class of
systems. Throughout, we will consider arbitrary quantum systems 
(or their  classical descendants)  in $D$- spatial (or space-time) dimensions 
with Hilbert spaces  which are  either finite or denumerably infinite. 
The theories to be directly analyzed  are defined on a $D$-dimensional 
lattice of {\it volume} $\Lambda$.  When alluded to, continuum field
theories will be understood to be represented by the  continuum
(vanishing lattice size $a$) limit of an appropriately defined lattice
field theory regularizing  it.   The Hamiltonians  (or
Euclidean actions) of 
the systems that we address have interactions {\it local}  in space (or
space-time), i.e., have couplings that veer to zero at large  spatial
(or temporal) separations.  
In the thermodynamic
($\Lambda \rightarrow \infty$) limit we will, at times, replace 
pertinent sums over the Fourier modes ${\bf k}$ associated with lattice
systems by  corresponding integrations over the first Brillouin zone.  For instance, in
the case of a square lattice system, $k_{\mu} = 2 \pi
l_{\mu}/L$ with $L$ the linear lattice size along the $\bm{e}_\mu$ direction, 
$\mu=1,2$,  $-L/2a <  l_{\mu} \le L/2a$ being integers, with $a$ the
lattice constant. The first Brillouin zone is spanned by $ -\pi/a <
k_\mu \le \pi/a$.

Symmetries represent an important element of our analysis. 
The most general symmetry \({\cal U}\) of a system with a Hamiltonian $H$ 
can be described as a direct sum  of arbitrary unitary transformations
$U(N_{E_{i}})$ that act non-trivially  only on their corresponding
subspaces of fixed energy $E_{i}$ and of  dimension $N_{E_{i}}$. That
is, the direct sum of unitary transformations
\begin{eqnarray}
{\cal U}= \sum_{{i}}\ U(N_{E_{i}})
\label{t}
\end{eqnarray}
constitutes a symmetry of the system,   i.e., ${\cal U}^{\dagger} H
{\cal U} = H$, for arbitrary $U(N_{E_{i}})$. Conversely, if \({\cal U}\)
is a symmetry and \(H=\sum_{{i}} E_i P_i\) is the spectral decomposition
of \(H\), then \({\cal U}\) has the structure of Eq. \eqref{t} above,
with \(U(N_{E_{i}})={\cal U}P_i\). Typically, unitary operations of the
type of Eq. (\ref{t}) may correspond to non-local operations. {\it In
contrast, the symmetries that are most relevant to dimensional reduction
have well defined spatial support}. 

This observation (that will be illustrated often in this paper) leads to
the notion of ${\bf d}$-GLSs. As alluded to earlier, 
these are symmetry operations (of codimension $D-{\bf d}$) that act
non-trivially only within a ${\bf d}$-dimensional  physical space. 
Technically, a symmetry is termed ${\bf d}$-GLS  if  fields acted on by
the symmetry operators ${\cal U}_{\bf d}$ are affected within a region
of $\bf d$-dimensional support \cite{NO,BN}.  For instance, local
(gauge) symmetries are of dimension ${\bf d}=0$. Similarly, global
symmetry operators act on all of the fields of the system and thus their
dimension is ${\bf d}=D$ (that of the entire system). In between these
two extremes of local (${\bf d}=0$) and global (${\bf d}=D$) symmetries
there is an intermediate regime where the symmetry operators act on a
$0<{\bf d}<D$ dimensional spatial region.  It was recently shown
\cite{NO} that  ${\bf d}$-GLSs can mandate the  appearance of
topological quantum order (TQO) \cite{wenbook}. 
In continuum space-time  ${\bf d}$-GLSs  may asymptotically include, as a
particular case,  the conformal symmetries that appear in  conformal
fields theories (CFTs) but are not limited to these.  

It will be useful in the following 
to keep in mind specific simple examples for which our
theorems have easy to grasp yet non-trivial consequences.
The spin $S=1/2$, $D=2$ compass model \cite{NO,BN} on a 
square lattice is one of many such examples. It is often specified by the Hamiltonian
\begin{eqnarray}
\label{compass}
H_{\sf compass}= - \sum_{\x \in \Lambda} ( J_1 \
\sigma^{x}_{\x} \sigma^{x}_{\x + \i} + J_2 \
\sigma^{y}_{\x} \sigma^{y}_{\x + \j}),
\end{eqnarray}
with exchange couplings $J_\mu$, 
 where $\sigma_{\x}^{\alpha=x,y}$ are Pauli operators located at 
the lattice site $\x=i^1\bm{e}_1+i^2\bm{e}_2$ ($\x \in \Lambda$).
Its \({\bf d}=1\)-GLSs 
\begin{equation}
\label{XYs}
X_{i^1}=\prod_{i^2}\sigma^x_{i^1,i^2},\ \ \ \ 
Y_{i^2}=\prod_{i^1} \sigma^y_{i^1,i^2},
\end{equation}
($[H_{\sf compass},X_{i^1}]=0=[H_{\sf compass},Y_{i^2}]$) satisfy the algebra
\begin{equation}
\ \{X_{i^1}, Y_{i^2}\}=0,\ \ \ \ [X_{i^1},X_{{i^1}'}]=[Y_{i^2},Y_{{i^2}'}]=0.
\label{algXY}
\end{equation}
The Xu-Moore model \cite{xu}, studied
before in connection to dimensional reduction, 
is dual to the compass model \cite{bondprl,NF}. 

\section{Effective dimensional reduction via bounds}
\label{Q}

As explained in the introduction, our approach to {\it effective  dimensional
reduction} is based on inequalities. These inequalities
are the subject of a theorem on an effective quantum dimensional reduction (EQDR)
that we prove in this section.  

Intuitively speaking, a system is intrinsically  \(D\)-dimensional if it cannot be
simulated by a lower-dimensional one (a higher-dimensional simulator
is always possible provided some redundancy is allowed \cite{somma}).
This idea provides the basis for a good definition of  {\it intrinsically
\(D\)-dimensional}, but fails to provide  simple, practical criteria to
search for dimensional reduction.  The EQDR theorem provides one such criterion.  
The idea is that \(D\)-dimensional systems have observables \(f\) that  
specifically probe \(d\)-dimensional sub-volumes \(\Gamma\), \(d\leq D\), of the
\(D\)-dimensional space \(\Lambda\). 
The EQDR theorem illustrates that under certain (not very stringent) conditions, 
the expectation values of these \(d\)-dimensional observables are both 
bounded from above and from 
below by expectation values taken within {\it effective}
\(d\)-dimensional theories. Thus, we can use the behavior of 
these bounds (e.g., how stringent they are for
specific theories) as indicators to search for effective dimensional
reduction. As stressed in the  introduction, the 
EQDR theorem  becomes most useful in the presence of suitable {\bf d}-GLSs.

The EQDR theorem exploits features
of quantum mechanics that force us to reconsider the  notion of {\it
localizability}. This notion poses a problem that was perhaps first
fully appreciated in the context of entanglement
\cite{einstein,schroedinger,GE} but was also carefully considered in 
connection to the notion of particle  (more specifically, the position of a particle) 
in quantum field theory \cite{newton}.  Remarkably, the notions of
localizability that emerged from those two lines of inquiry are (as far
as we can see) conceptually different, and exploit different 
mathematical structures on the space of states.
Hence, in order to apply the EQDR theorem to quantum field theories and many-body systems
we need to introduce a redundant representation of the state space of 
identical particles, and keep the constraint of indistinguishability explicit
in the form of a projector onto subspaces of completely (anti)symmetric state 
vectors. This enables us to treat fermions and bosons. In the current work, we do not discuss 
the extension of the EQDR theorem to systems with anyon statistics.

\subsection{A theorem on effective quantum dimensional reduction (EQDR theorem)}
\label{qdrtheorem}

The EQDR theorem is inspired by an analogous theorem for classical 
statistical mechanics \cite{BN}. Hence it is  convenient to review 
briefly this theorem on classical dimensional reduction to motivate its 
quantum analogues and to introduce some useful notation.

Consider a {\it classical} model of statistical mechanics that, for
concreteness, has its degrees of freedom defined on the sites of a
lattice \(\Lambda=\Gamma\cup \bar{\Lambda}\). The lattice provides a
{\it set of labels}. In the following,  we will loosely allude to \(\Gamma\)
as ``the boundary'' of the system 
and to \(\bar{\Lambda}\) as ``the bulk'' of the system. The  elementary degrees of
freedom of the model can be organized into a lattice field
\({\phi}(\x)\) (\(\x \in \Lambda\) denotes a site of the lattice) that
can be broken up into two fields with disjoint supports,  $\Gamma \cap
\bar{\Lambda}=\emptyset$, 
\begin{equation}
\phi(\x)=\left\{
\begin{array}{rcl}
{\phi}_{0}(\x) & \mbox{if} & \x\in\Gamma \\
\psi(\x) & \mbox{if} & \x \in \bar{\Lambda}
\end{array}
\right. \ .
\end{equation}

If \(f\) is a classical observable with support on \(\Gamma\) then we
can write \(f[\phi]=f[\phi_{0}]\) so that its canonical ensemble
average is
\begin{eqnarray}\label{splitclassave}
\langle f\rangle^D=\sum_{\{\psi\}}\sum_{\{\phi_0\}} 
f[\phi_0]\frac{e^{-\beta E[\phi_0,\psi]}}{\mathcal{Z}}
= \sum_{\{\psi\}}\  \frac{z[\psi]}{\mathcal{Z}}\ \frac{
\sum_{\{\phi_0\}}f(\phi_0)e^{-\beta E[\phi_0,\psi]}}{z[\psi]}\ ,
\end{eqnarray}
where $E[\phi]=E[\phi_0,\psi]$ is the energy functional, \(\mathcal{Z}=\sum_{\{\phi\}}
e^{-\beta E[\phi]}\), with  $\beta=1/k_B T$ where
$k_B$ is Boltzmann's constant, and  \(z[\psi]={\sum_{\{\phi_0\}}e^{-\beta
E[\phi_0,\psi]}}\). The quantity
\begin{equation}
\langle f\rangle^d[\psi]\equiv\frac{
\sum_{\{\phi_0\}}f(\phi_0)e^{-\beta E[\phi_0,\psi]}}
{z[\psi]}
\end{equation}
represents a {\it conditional} expectation value of $f$ dependent on the
condition that within the bulk of the system
$\bar{\Lambda}$ the field assumes the value $\psi$.  
The probability distribution associated with this condition is
\(p(\psi)=z[\psi]/\mathcal{Z}\). Let us furthemore define
\begin{equation}
\langle f\rangle^d_l \equiv {\sf min}_{\psi}\ \langle f\rangle^d[\psi] =
\langle f\rangle^d[\psi_{\sf min}], \ \ \ \ \ \ 
\langle f\rangle^d_u \equiv {\sf max}_{\psi}\ \langle f\rangle^d[\psi] =
\langle f\rangle^d[\psi_{\sf max}].
\end{equation}
It then follows from Eq. \eqref{splitclassave} that \cite{BN}
\begin{equation}\label{cinq}
\langle f\rangle^d_l\leq \langle f\rangle^D\leq\langle f\rangle^d_u.
\end{equation}

The key observation is that we can think of these bounds as averages 
for effective \(d\)-dimensional theories. Moreover, these theories are characterized by {\it
local} energy functionals
\begin{equation}
E_l[\phi_0,\psi_{\sf min}] \ \ \ \ \mbox{and}\ \ \ \ 
E_u[\phi_0,\psi_{\sf max}].
\end{equation}
Thus, the inequalities of Eq. \eqref{cinq} afford a classical notion of
{\it effective} dimensional reduction. For some systems, it is, to a certain extent,
a matter of taste whether we choose to call these inequalities ``classical'' 
or ``quantum'' as they are directly applicable to {\it lattice} quantum  field
theory in the Euclidean path-integral representation \cite{LQFT}.  
However, the path-integral formalism is not always convenient,
and some subtleties may arise when this formalism is applied to quantum theories with
a boundary. Thus we would like to understand next how to extend this type of reasoning 
to the operator formalism of quantum mechanics. 

Suppose that we have a quantum system occupying a
\(D\)-dimensional spatial volume \(\Lambda\). This system is described by the 
density matrix \(\rho\) that we take to be completely arbitrary for
now (in particular, it could be time dependent). 
That is, the density matrix
$\rho$ is solely required to be a positive and 
normalized \(\tr_\Lambda(\rho)=1\) operator acting on the  state space
\(\mathcal{H}_{\Lambda}\), with \(\tr_\Lambda\)  denoting the trace on
\(\mathcal{H}_\Lambda\).  A natural question to ask is 
whether \(\rho\) should play in the quantum arena 
a role similar to that played by the Boltzmann probability
distribution in the classical setting.  
Another important question is ``what does it mean for a quantum
observable \(f\) to be localized on  \(\Gamma\)?''. The answer to the latter question depends on
how the {\it set} decomposition \(\Lambda=\Gamma \cup \bar{\Lambda}\) of the labels
is reflected in the state space \(\mathcal{H}_{\Lambda}\). If, 
{\it for whatever physical reason}, the Hilbert space associated with 
the entire system can be expressed as 
a direct product of the spaces associated with the boundary and bulk,
\begin{equation}\label{distbb}
\mathcal{H}_{\Lambda}=\mathcal{H}_{\Gamma}\otimes\mathcal{H}_{\bar{\Lambda}}\ ,
\end{equation} 
i.e., {\it if the boundary $\Gamma$ and bulk $\bar{\Lambda}$ 
are distinguishable},  then we can use the
standard notion of subsystems in quantum mechanics to say that \(f\) {\it
is localized on \(\Gamma\)} provided that
\begin{equation}
f=f_\Gamma \otimes \mathds{1}_{\bar{\Lambda}}.
\end{equation}
Once we agree on this notion of localizability, the EQDR theorem
follows from a simple direct computation involving partial traces.

First we notice that
\begin{eqnarray}
\langle f\rangle^D=\tr_\Lambda(\rho
f)=\tr_{\bar{\Lambda}}\tr_\Gamma(\rho f),
\label{splitq}
\end{eqnarray}
where  \(\tr_\Gamma\) denotes the {\it partial} trace operation that
takes an operator on 
\(\mathcal{H}_\Lambda=\mathcal{H}_\Gamma\otimes\mathcal{H}_{\bar{\Lambda}}\)
to an operator on the state space for the bulk, 
and \(\tr_{\bar{\Lambda}}\) denotes the standard trace on \(\mathcal{H}_{\bar{\Lambda}}\). 
Equation \eqref{splitq} is analogous to the
first part of Eq. \eqref{splitclassave} in the classical case. To reproduce the second
part, {\it we assume that the reduced
operator \(\tr_\Gamma(\rho)\) is invertible}, so that we can write 
\begin{equation}\label{qbcond}
\langle f\rangle^D=\tr_{\bar{\Lambda}}\Big(\tr_\Gamma(\rho)\big[
\tr_\Gamma(\rho)^{-1}\tr_\Gamma(\rho f)\big]\Big).
\end{equation}
Very loosely speaking, in such a case, we can think of
\(\tr_\Gamma(\rho)^{-1}\tr_\Gamma(\rho f)\) as a ``conditional 
probability"  where the conditions pertain to the quantum states of the
bulk. Explicitly, the reduced operator \(\tr_\Gamma(\rho)\) can be decomposed
in terms  of pure states of the bulk as  \(\tr_\Gamma(\rho)=\sum_i\
r_{\bar{\Lambda}i} P_{\bar{\Lambda}i}\), where \(P_{\bar{\Lambda}i}\) is
the projector onto the eigenspace with the eigenvalue  \(r_{\bar{\Lambda}i}\).
This (positive) eigenvalue represents the probability to find the bulk
in the state \(P_{\bar{\Lambda}i}\). Then,
\begin{equation}\label{Dqave}
\langle f\rangle^D=\sum_i\ r_{\bar{\Lambda}i}\
\tr_{\bar{\Lambda}}\big(P_{\bar{\Lambda}i}
\big[\tr_\Gamma(\rho)^{-1}\tr_\Gamma(\rho f)\big]\big).
\end{equation}

To compare Eq. \eqref{Dqave} with its counterpart
Eq. \eqref{splitclassave} in the classical dimensional reduction theorem,  
notice that Eq. \eqref{splitclassave} expresses
the classical average as a conditional expectation value
weighted by the probabilities for different possible bulk
configurations. Equation \eqref{Dqave} has a similar structure. However,
while the left-hand side of Eq. \eqref{Dqave} is real by construction, 
the quantities \(\tr_{\bar{\Lambda}}\big(P_{\bar{\Lambda}i}
\big[\tr_\Gamma(\rho)^{-1}\tr_\Gamma(\rho f)\big]\big)\) need not be
real valued since the operator \(\tr_\Gamma(\rho)^{-1}\tr_\Gamma(\rho
f)\)  need not, in general, be Hermitian. [This holds even if we assume that \(f\) is
Hermitian (an assumption that we will not invoke)]. Later on, it will be
useful to consider the situation when \(f\) is normal (\([f,f^\dagger]=0\)). Thus, we
should think of   $\tr_{\bar{\Lambda}}\big(P_{\bar{\Lambda}i}
\big[\tr_\Gamma(\rho)^{-1}\tr_\Gamma(\rho f)\big]\big)$ as a generalized
conditional average, in the spirit for instance of Dirac's work 
\cite{dirac}.  With this understanding we define
\begin{eqnarray}
\langle f\rangle^d_l&\equiv&{\sf min}_i\ {\sf Re}\Big \{
\tr_{\bar{\Lambda}}\big(P_{\bar{\Lambda}i}
\big[\tr_\Gamma(\rho)^{-1}\tr_\Gamma(\rho f)\big]\big) \Big \}= \langle
f\rangle^d_{i_{\sf min}},\label{qb1}\\
\langle f\rangle^d_u&\equiv&{\sf max}_i\ {\sf Re}\Big
\{\tr_{\bar{\Lambda}}\big(P_{\bar{\Lambda}i}
\big[\tr_\Gamma(\rho)^{-1}\tr_\Gamma(\rho f)\big]\big) \Big \}= \langle
f\rangle^d_{i_{\sf max}}.
\label{qb2},
\end{eqnarray}
Thus, from Eq. \eqref{Dqave} it follows that
\begin{equation}\label{strong1}
\langle f\rangle^d_l\leq \langle f\rangle^D\leq\langle f\rangle^d_u\ .
\end{equation}

Thus
we have, in Eq. (\ref{strong1}), proved the first theorem of this 
paper:

\vspace*{0.2cm}

\noindent
{\bf Theorem 1 (EQDR).} 
Consider a quantum system in the state \(\rho\), occupying a
\(D\)-dimensional region \(\Lambda\), with a decompositon of \(\Lambda\) 
into a \(d\)-dimensional boundary \(\Gamma\) and its complementary bulk 
\(\bar{\Lambda}\), such that $\Lambda= \Gamma \cup \bar{\Lambda}$. 
Suppose that the state space of the system admits a  tensor product
decomposition \(\mathcal{H}_\Lambda=
\mathcal{H}_\Gamma\otimes\mathcal{H}_{\bar{\Lambda}}\), and let 
\(f=f_\Gamma\otimes\mathds{1}_{\bar{\Lambda}}\) be localized on
\(\Gamma\). If the subsystem density matrix \(\tr_{\Gamma}(\rho)\) of the bulk 
is invertible, and if the $D$-dimensional system is local then the expectation value \(\langle
f\rangle^D=\tr_\Lambda(\rho f)\) can be bounded both from below and from
above (Eq. \eqref{strong1}) by suitably  defined expectation values (Eqs.
\eqref{qb1} and \eqref{qb2}) of \(f_\Gamma\) for {\it local}
$d$-dimensional ($d<D$) systems.  
\vspace*{0.2cm}

Some comments are in order. First, the {\it range} of the interactions in the
effective theories providing the bounds {\it is the same} as in the original
\(D\)-dimensional theory. Thus, as alluded to above, if the high ($D$)-dimensional system  is 
local then no long-range effective interactions will appear in
the low ($d$)-dimensional system. This is different from what would
generally be expected if a lower-dimensional description were to be achieved  by
brute force integration of bulk fields (in $D$ dimensions) in order to generate a
lower ($d$)-dimensional boundary theory. Second, we
can gain some insight into the condition  that \(\tr_{\Gamma}(\rho)\) be
invertible by noticing, for example, that a finite temperature density matrix corresponding to
a system Hamiltonian $H$,  i.e., \(\rho=e^{-\beta H}/\tr_\Lambda (e^{-\beta
H})\), is always invertible. The same applies for the corresponding reduced 
\(\tr_{\Gamma}(\rho)\). It is interesting to notice, however, that the 
Boltzmann distribution cannot be (easily)
extended to gravitational systems \cite{explain_gravity_problems} already in their classical
incarnation. Third, there is a different version
of the EQDR theorem that holds for arbitrary 
\(\rho\), and that we expand in \ref{alternativeeqdr}.
Fourth, while we choose to refer to \(\Lambda\) as the volume occupied by
the system, \(\Lambda\) could, in fact, represent any suitable {\it set of
labels or modes} without reference to any specific notion of space (or space-time). For
example, if \(\Lambda\) denotes momentum space then our
theorem will refer to {\it localization} and {\it dimensional
reduction} in momentum space. Fifth, the EQDR theorem puts no restrictions
on \(f\) other than \(f=f_\Gamma\otimes \mathds{1}_{\bar{\Lambda}}\). In particular,
\(f\) does not have to be Hermitian. Finally, with an eye towards applications
of the inequalities, we remark that  we  may choose $f$ to be invariant 
with respect to a given group of ${\bf d}$-GLSs
$\{{\cal U}_{\bf d}\}$.  
That is, we may replace $f$  by an operator 
$f_{\sf sym}$ which is  symmetrized with respect to $\{{\cal U}_{\bf d}\}$,
\begin{eqnarray}
f \mapsto f_{\sf sym}\equiv \sum_{\{{\cal U}_{\bf d}\}}~ {\cal U}_{\bf
d} \, f \, {\cal U}_{\bf d}^{\dagger}\ .
\label{symf}
\end{eqnarray}

\subsubsection{The EQDR theorem in quantum field theory and 
many-body physics}

The natural state space in quantum field 
theory and many-body physics is that of Fock space which automatically takes into
 account 
the quantum statistics of {\it indistinguishable} particles. A Fock space can be
described most efficiently as an irreducible representation of an algebra
of operators of {\it particle} creation (\(d_{\x_i}^\dagger\)) and annihilation 
(\(d^{\;}_{\x_i}\))
on a vacuum state vector.
The actual kind of statistics is not important in what follows, so we consider
operators identified by some suitable {\it
set of labels} that we denote generically by \(\x_i, \x_j\in\Lambda\),  and that
satisfy
\begin{equation}\label{algqstat}
[d^{\;}_{\x_i},d_{\x_j}^\dagger]_{\pm}=\delta_{\x_i,\x_j}, \ \ \ \
[d^{\;}_{\x_i},d^{\;}_{\x_j}]_\pm=0,\ \ \ \ [d_{\x_i}^\dagger,d_{\x_j}^\dagger]_\pm=0,
\end{equation}
where \([A,B]_\pm \equiv AB\pm BA\). These operators act on a Fock space 
\begin{equation}\label{fock}
\mathcal{H}^{\sf Fock}=\mathcal{H}_0\oplus \mathcal{H}_1
\oplus \mathcal{H}_2\oplus \cdots=\bigoplus_{n=0}^\infty \mathcal{H}_n,
\end{equation}
where \(\mathcal{H}_0\) is the subspace spanned by  the
vacuum state \(|0\rangle\), and \(\mathcal{H}_n\) is spanned by the
$n$-particle states
\begin{equation}
d_{\x_1}^\dagger d_{\x_2}^\dagger \cdots d^\dagger_{\x_n}|0\rangle\  
\end{equation}
(\(\x_i,\ i=1,\cdots, n\), need not be all different if we are  dealing
with bosons).

The explicit description of the Fock space shows
that there is a difficulty in applying the EQDR theorem to quantum field theory
directly, since in this case the decomposition \(\Lambda=\Gamma\cup \bar{\Lambda}\)
of the set of labels into bulk and boundary is not readily reflected in
the decomposition of \(\mathcal{H}^{\sf Fock}\) into a tensor product as in 
Eq. \eqref{distbb}. To find a solution to this problem, let us imagine
for a moment that the particles in the bulk $\bar{\Lambda}$ are {\it distinguishable} from 
the particles in the boundary $\Gamma$, keeping however the particles in the
boundary (bulk) indistinguishable. Then the space \(\mathcal{H}_\Lambda\)
becomes
\begin{equation}\label{lfock}
\mathcal{H}_\Lambda=\mathcal{H}^{\sf Fock}_\Gamma\otimes\mathcal{H}^{\sf Fock}_{\bar{\Lambda}}
=\left(\bigoplus_{n=0}^{\infty} \mathcal{H}_{\Gamma n}\right)\otimes 
\left(\bigoplus_{n=0}^{\infty} \mathcal{H}_{\bar{\Lambda}n}\right).
\end{equation}
This space has the right properties to make our EQDR theorem applicable. 
 The true (irreducible) state space of the system, the Fock
space $\mathcal{H}^{\sf Fock}$, can be realized as a proper subspace of \(\mathcal{H}_\Lambda\).
To see this, notice that we can reorganize the right-hand side of Eq. \eqref{lfock} as
\begin{equation}
\mathcal{H}_\Lambda=\bigoplus_{n=0}^{\infty}\ \bigoplus_{m+m'=n} 
{\cal H}_{\Gamma m}\otimes {\cal H}_{\bar{\Lambda}m'},
\end{equation}
thus decomposing  \(\mathcal{H}_\Lambda\) into subspaces with fixed numbers of particles.
These subspaces include \(n\)-particle states that do not respect the indistiguishability
of the particles in the whole system, but contain  as a subspace
\begin{equation}
\mathcal{H}_n\subset \bigoplus_{m+m'=n} H_{\Gamma m}\otimes H_{\bar{\Lambda}m'},
\end{equation}
the space of {\it properly (anti-)symmetrized} \(n\)-particle states. 

We can then define an orthogonal projector \(P_{\sf Fock}\) such that 
\begin{equation}
P_{\sf Fock}\mathcal{H}_\Lambda P_{\sf Fock}=\mathcal{H}^{\sf Fock},
\end{equation}
and describe the system in terms of \(\mathcal{H}_\Lambda=
\mathcal{H}^{\sf Fock}_\Gamma\otimes\mathcal{H}^{\sf Fock}_{\bar{\Lambda}}\)
{\it and a superselection rule}: a positive, normalized  operator \(\rho\) 
represents a quantum state {\it if and only if}
\begin{equation}\label{stateind}
\rho=P_{\sf Fock}\rho P_{\sf Fock}. 
\end{equation}  

Put differently, in this formulation of the EQDR theorem, the state 
{\it carries the statistics}.
Positive, normalized operators that do not satisfy Eq. \eqref{stateind} do not respect
the fact that we are dealing with a system of indistinguishable particles, and 
must be discarded \cite{ssobs}.  
This reformulation of the kinematics of quantum fields/many-body problems makes the
application of the EQDR theorem straightforward: everything works as in Section 
\ref{qdrtheorem},
whenever the state \(\rho\) satisfies the  physical condition of Eq. \eqref{stateind}.

\subsubsection{A basis-dependent version of the EQDR theorem}

The proof of the EQDR theorem did not rely on the use of any particular basis
of ${\cal H}_\Lambda$.  Here we explore the consequences of a
basis-dependent manipulation that we will later use in some examples,
and helps to emphasize the connections to Hamiltonian lattice quantum
field theory.

As in Refs. \cite{NO,BN}, we may consider arbitrary functions $f$   of
the boundary fields $\tilde{\phi}_{0}(\x)$ and its conjugate momentum 
$\tilde{\pi}_{0}(\x) = -i \delta/\delta \tilde{\phi}_0(\x)$.  In such a
case, $f$ acts trivially on the  bulk fields $\psi(\x)$. Consider  the
orthonormal basis $\{|\phi_0 \rangle\}$  (without a tilde) to be an
eigenbasis of $f$ on the boundary $\Gamma$ (meaning that $f$ is a 
normal operator).  The product states $\{|\phi_0 \rangle \otimes |\psi
\rangle= |\phi_0 \psi\rangle\}$, i.e., a direct product basis of states
of $\Gamma$ and the bulk $\bar{\Lambda}$, form an orthonormal basis on
which all physical states $\{ |\phi \rangle \} \in {\cal H}_\Lambda$ can
be spanned, and are, trivially, eigenstates of $f$ as well.   For
instance, for a situation  in which $f =
\tilde{\pi}_{0}(\x_1)\tilde{\pi}_{0}(\x_2)$, we will choose the basis to
be spanned by the set  of product states, such that  $f |\phi_0
\psi\rangle=\tilde{\pi}_{0}(\x_1)\tilde{\pi}_{0}(\x_2) |\phi_0
\psi\rangle$. 
 
We need to evaluate $\tr_\Lambda(\rho f)=\tr_{\{\phi_o\psi\}}(\rho f)$.
A  general matrix element 
\begin{eqnarray}
\langle \phi_0 \psi| \rho f| \phi_0 \psi \rangle \! = \! \!\!\!\!
\sum_{\{|\phi_0' \psi' \rangle\}} \! \!\! \langle \phi_0 \psi| \rho
|\phi_0' \psi' \rangle \langle  \phi_0' \psi' |f | \phi_0 \psi \rangle
\! = \! \langle \phi_0 \psi| \rho |\phi_0 \psi \rangle  \langle \phi_0 
\psi| f | \phi_0 \psi \rangle, 
\label{sep}
\end{eqnarray}
as $\{ | \phi_0 \psi \rangle \}$ are eigenstates of $f$. Thus,
\begin{eqnarray}
\langle f\rangle^D=\tr_{\{\phi_o\psi\}}(\rho
f)=\tr_{\{\phi_o\psi\}}(\rho_{\sf diagonal} f),
\end{eqnarray}
where the density operator $\rho_{\sf diagonal}$  is defined as
\begin{eqnarray}
\rho_{\sf diagonal} &\equiv& \sum_{\{|\phi_{0}  \psi \rangle\}} |
\phi_{0} \psi \rangle  \langle \phi_{0}  \psi| \rho| \phi_{0} \psi
\rangle \langle  \phi_{0} \psi| ,
\end{eqnarray}
such that $\langle \phi_{0} \psi| \rho_{\sf diagonal}|\phi_{0} \psi
\rangle \ge 0$,  reflecting the fact that only the diagonal elements of
$\rho$ matter  in such a case.  If we define a reduced density operator on
$\bar{\Lambda}$ 
\begin{eqnarray}
\rho_{\sf diagonal}[\psi] \equiv \tr_{\{\phi_0\}} (\rho_{\sf
diagonal}),
\end{eqnarray}
then, clearly,
\begin{eqnarray}
 \tr_{\{\psi\}}(\rho_{\sf diagonal}[\psi])
=\tr_{\{\phi_{0}\psi\}}(\rho_{\sf diagonal})
=\tr_{\{\phi_{0}\psi\}}(\rho)=1.
\label{fund}
\end{eqnarray}
The average of $f$ within the complete volume $\Lambda$
(containing both the boundary and the bulk), 
\begin{eqnarray}
\langle f \rangle^D = \tr_{\{\psi\}} (\tr_{\{\phi_{0}\}}( \rho_{\sf
diagonal} f) )  = \tr_{\{\psi\}} ( \rho_{\sf diagonal}[\psi] \langle f 
\rangle^d_{\psi}) ,
\label{Z}
\end{eqnarray}
is the expectation value of $f$ with fixed fields $\psi$ in the bulk
$\bar{\Lambda}$.  In the last equality of Eq. (\ref{Z}) 
we used the fact that $\rho_{\sf
diagonal}[\psi]$ is invertible,
\begin{eqnarray}
\langle  f \rangle^d_{\psi} \equiv \rho_{\sf diagonal}[\psi]^{-1}
 \tr_{\{\phi_{0}\}}( \rho_{\sf diagonal} f) .
\label{Bayes}
\end{eqnarray}

We are interested in establishing bounds on $\langle  f \rangle^D$. 
The latter  may represent an  arbitrary correlation function on the
$d$-dimensional sub-volume $\Gamma$. Proceeding as
in earlier subsections, we obtain
\begin{eqnarray}
\langle  f \rangle^d_{\psi_{\min}} \le  \langle f  \rangle^D  \le 
\langle f \rangle^d_{\psi_{\max}} ,
\label{strong}
\end{eqnarray}
where  $\{ \psi_{\max} \}$  and $\{\psi_{\min}\}$ represent  the
collection of fields in the bulk for which  $\langle f \rangle^d_{\psi}$
is maximal/minimal. The expectation values $\langle f \rangle^d_{\psi}$
are weighted with a normalized positive weight $\rho_{\sf
diagonal}[\psi]$.  With the substitution of
these fields,  Hamiltonians such as  $H_{\psi_{\max}}[\tilde{\phi}_{0}]$
and $H_{\psi_{\min}}[\tilde{\phi}_{0}]$ (or corresponding actions) will
be functionals of only the boundary fields $\tilde{\phi}_{0}(\x)$. Furthermore, 
similar to what we re-iterated earlier (and do so again now), 
these Hamiltonians will be {\em local} in $d$  dimensions (i.e., involve only
local interactions) if the original $D$-dimensional theory is local.  As
further noted earlier, this locality is to be contrasted with a brute force
integration of the bulk fields $\psi$ to obtain {\em exact} effective
Hamiltonians $H_{\sf eff}[\tilde{\phi}_{0}]$ within the sub-volume
$\Gamma$, where arbitrary long-range interactions may  be involved.

We note that if we choose the quantity $f$ in Eq. (\ref{strong}) to be the
symmetrized function $f_{\sf sym}$ of Eq. (\ref{symf}) then the
minimizing fields $\psi_{\min}$ (and  similarly the maximizing fields
$\psi_{\max}$) will be related to one another by the symmetry operations
($\bf d$-GLSs) ${\cal U}_{\bf d}$. In particular, if  $\psi_{\min}$ (or  $\psi_{\max}$)
transform as a singlet under the symmetry operations ${\cal U}_{\bf d}$
then  the averages $\langle f_{\sf sym} \rangle^{\bf
d}_{\psi_{\max,\min}}$ are computed with an effective local $\bf
d$-dimensional  Hamiltonian (action) that respects the symmetry ${\cal
U}_{\bf d}$. 

\subsection{The role of symmetries}
\label{ros}

We next consider the consequences of symmetries on the bounds of Eq.
(\ref{strong}) for systems for which the 
(correlation) functions $f$, the pertinent projection operators, and
symmetry generators that have their support on  different spatial
regions commute with one another (and thus Eqs. (\ref{coleman1},
\ref{coleman2})  below will be satisfied).  We note that with the fixed
bulk fields $\psi_{\max}$ and $\psi_{\min}$ that form the bounds in Eq.
(\ref{strong}), the $d$-dimensional Hamiltonians
$H_{\psi_{\max}}[\tilde{\phi}_{0}]$ and
$H_{\psi_{\min}}[\tilde{\phi}_{0}]$ (or actions) will admit a {\it
global} ${\bf d}$-GLS symmetry on the region $\Gamma$ if the original
$D$-dimensional theory exhibits a ${\bf d}$-GLS on that region. (This is
so as by virtue of the existence of ${\bf d}$-GLSs, the bulk fields
$\psi$  must act as external non-symmetry breaking fields in
$H_{\psi_{\max}}[\tilde{\phi}_{0}]$  and
$H_{\psi_{\min}}[\tilde{\phi}_{0}]$ or any other Hamiltonian formed by
fixing $\psi$:  $H_{\psi}[\tilde{\phi}_{0}]$.)  The inequalities in Eq.
(\ref{strong}), leading to the concept of dimensional  reduction in the
{\it weak sense} of upper and lower bounds,  are quite general for
systems displaying ${\bf d}$-GLSs.  As noted earlier, the operators $f$
are quite arbitrary and may be chosen to  be arbitrary $n$-point
correlation functions. When $\{\x_{i}\}_{i=1}^{n} \in \Gamma$, the exact
$n$-point correlators $G^{(n)}$ within the $D$-dimensional theory are
bounded, both from above and from below, by correlation functions in
{\em local theories} in $\bf d$ dimensions (theories defined on $\Gamma$
alone) {\em that exhibit ${\bf d}$-GLSs},
\begin{eqnarray}
G^{(n)}_{{\sf lower}; {\bf d}}(\{\x_{i}\}) \le G^{(n)}_{{\sf exact};
D}(\{\x_{i}\})   \le G^{(n)}_{{\sf upper};{\bf d}} (\{\x_{i}\}).
\label{g_n}
\end{eqnarray}
Equation (\ref{g_n}) fleshes out a corollary of the EQDR theorem for
systems with ${\bf d}$-GLSs. Similar to our discussion of Eq.
(\ref{symf}), we may choose the correlators $G^{(n)}$ to be symmetrized
relative to the symmetry generators. This may effectively set  the
averages $\langle G^{(n)} \rangle_{{\sf upper/lower};{\bf d}}$ to be
those computed with a local $\bf d$-dimensional theory that respects
the  symmetry generators at hand.

It is important to emphasize the reason for being able to freeze out the
bulk degrees of freedom in obtaining the inequality $(\ref{strong})$,
while maintaining  ${\bf d}$-GLSs in the bounding theories if the
original theory exhibits ${\bf d}$-GLSs.  Let us define the projection operator
$P_{\psi}$ which has
its support only on degrees of freedom in the bulk
$\bar{\Lambda}$, while the lower-dimensional symmetries  $\{{\cal U}_{\bf d}\}$
have their support on fields in the sub-volume $\Gamma$. Therefore,
\begin{eqnarray}
\label{coleman1}
[P_{\psi}, {\cal U}_{\bf d}]=0
\end{eqnarray}
as $\bar{\Lambda} \cap  \Gamma = \{
\emptyset\}$. As ${\cal U}_{\bf d}$ are symmetries of $\rho$, we then have that 
\begin{eqnarray}
[P_{\psi} \rho P_{\psi}, {\cal U}_{\bf d}]= P_{\psi} [\rho,{\cal U}_{\bf d}] P_{\psi}
= 0.
\end{eqnarray}

The expectation value of Eq. (\ref{Bayes}) is that of 
$f({\tilde{\phi}}_{0})$ with the density operator $\rho$ computed
over the projected density operator $\rho_{\psi} \equiv P_{\psi} \rho
P_{\psi}$ when it is invariant  under ${\cal U}_{\bf d}$.  This is so since 
\begin{eqnarray}
[f({\tilde{\phi}}_{0}), P_{\psi}]=0
\label{coleman2}
\end{eqnarray}
(as, once again, the two operators
$f$ and $P_{\psi}$ have their  support on $\Gamma$ and 
$\bar{\Lambda}$, respectively). This commutation relation implies that, e.g., in the canonical
ensemble, 
\begin{eqnarray}
\langle  f \rangle^{\bf d}_{\psi} &=& \frac{\tr_{\{\phi_{0}' \psi'\}} (
P_{\psi} f \ e^{-\beta H} P_{\psi})}{\tr_{\{\phi_{0}' \psi'\}} (
P_{\psi} e^{-\beta H} P_{\psi})}  = \frac{\tr_{\{\phi_{0}' \psi'\}}(f
e^{-\beta H_{\psi}})} {\tr_{\{\phi_{0}' \psi'\}}(e^{-\beta H_{\psi}})}.
\end{eqnarray}

It is interesting to highlight those cases in which  $f$ is not
invariant under low-dimensional GLSs  (${\bf d}=0,1$ discrete symmetries
or ${\bf d}=0,1,2$ continuous symmetries) \cite{NO} in a system with
interactions of finite range and strength. Then, as such symmetries
cannot be broken in the low $\bf d$-dimensional system, the expectation
value of $f$ must vanish at finite temperatures  \cite{NO,BN}.  The case
of ${\bf d}=0$ corresponds to {\it Elitzur's theorem} for usual gauge theories
(with local (${\bf d}=0$) symmetries)  \cite{Elitzur}. One-dimensional
systems with interactions of finite range and strength do not exhibit 
spontaneous symmetry breaking at non-zero temperatures. Similarly,
two-dimensional systems do not exhibit spontaneous symmetry  breaking of
continuous symmetries. Thus, Elitzur's theorem can be extended to such 
cases \cite{NO,BN}. Similarly, if a low $\bf d$-dimensional correlation
function decays (exponentially or algebraically) with distance then by
virtue  of our inequalities the related correlation function in the
$D$-dimensional theory is similarly bounded. The physical engine behind
the absence of symmetry breaking  (as well as more exotic
high-dimensional {\it fractionalization}) in high-dimensional systems
with these symmetries  is that the energy-entropy balance in these 
$D$-dimensional systems is essentially that of ${\bf d}<D$ dimensional
systems \cite{NO,BN,NBF}. 

Our EQDR theorem and the ensuing consequences of  $\bf d$-GLSs may
seem rather formal. To elucidate the physical consequences of these symmetries, we turn to the
example of the square lattice compass model of Eq. (\ref{compass})
\cite{NO,BN}.  Let us first consider the consequences of the EQDR theorem of
Eq. (\ref{strong}) for this system. If we consider  the boundary
$\Gamma$ to be a line (i.e., a $d=1$ dimensional object) along a lattice
direction (say, the $\i$ direction) in the full two dimensional $(D=2$)
system of $H_{\sf compass}$,  then the
correlation function $\langle f \rangle^D= \langle \sigma^{x}_{\x}
\sigma^{x}_{\x'} \rangle$  between two sites $\x$ and $\x'$ that lie on
the line $\Gamma$ is bounded (both from above and from below) by the
same correlation function  in one-dimensional systems. These
one-dimensional systems in which we compute $\langle f
\rangle^{\bf d}_{\psi_{\max}}$ and $\langle f \rangle^{\bf d}_{\psi_{\min}}$ may be seen
to have an effective Hamiltonian that is composed of nearest-neighbor
Ising type interactions along $\Gamma$ augmented by external transverse
fields (originating from the  $\alpha =y$ contributions in Eq.
(\ref{compass}) once we fix the bulk fields (spins in this case) on all
sites $\y \in \bar{\Lambda}$, i.e.,  outside the line $\Gamma$, to a set
of particular values $\{\psi_{\min}\}$ or $\{\psi_{\max}\}$ (or
$\{\sigma^{y}_{\y; \min}\}, \{\sigma^{y}_{\y};_{\max}\}$ in this case). 
That is, the averages, as in Eqs. (\ref{qb1}, \ref{qb2}), are performed for $d=1$
systems. For each vertical or horizontal line $\Gamma$ we have 
the symmetries of Eq. (\ref{XYs}).   On any given $d=1$ line $\Gamma$, these 
symmetries will appear as global (${\bf d}=1$) symmetries. 
The external (transverse) fields $\psi$ do not spoil these symmetries.  

As in any such transverse field type Ising chain, there is no long-range
order at any finite temperature and coupling, 
the correlation function of $\langle
\sigma^{x}_{\x} \sigma^{x}_{\x'} \rangle$  tends to zero as sites $\x$
and $\x'$ become well separated. 
As noted in Eq. \eqref{XYs}, $H_{\sf compass}$ is invariant under the
following ${\bf d} =1$ (Ising type) symmetry operations of ${\cal U}_{\bf d}=
\prod_{\x \in \tilde{\Gamma}} \sigma_{\x}^{y}$ where  $\tilde{\Gamma}$
denotes any (${\bf d}=1$) line parallel to the $\i$-axis . (A similar set
of symmetries applies for lines parallel to the $\j$-axis with the  $x$
and $y$ Pauli matrices interchanged.) In this case, we may, in
particular, choose  $\Gamma$ to coincide with the region
$\tilde{\Gamma}$ where the symmetry operates.  On the line $\Gamma$, we
have (as stated earlier) in the bounding theories on both the left and
right-hand sides of Eq. (\ref{strong}),  a short-range one-dimensional
Ising model compounded by transverse fields which do not break the ${\bf
d}=1$ Ising symmetry above.  
This absence of symmetry breaking constitutes an analogue and extension 
\cite{NO,BN} of Elitzur's theorem that pertains to local gauge (i.e.,
${\bf d}=0$) symmetries \cite{Elitzur}.  When $J_1 \neq J_2$, this
analysis implies that the $D=2$ compass model does not display
long-range order at any finite temperature. At the isotropic $J_1=J_2$
point an additional ${\bf d}=D=2$ discrete symmetry appears which  can
be broken.

\section{Exact dimensional reduction: general scope}

In the next two sections, we discuss two related forms of exact dimensional
reduction. The two cases studied are those that appear in (a) various
limits such as large $n$ in $O(n)$ vector theories or high temperature
(or weak coupling) and (b) dimensional reductions via  exact dualities
that do not appear in some limit of the theory but can rather be
established by considering the relevant degrees of freedom (so called
{\it bonds})   that define the Hamiltonian  (or action) of the theory
\cite{bondprl,bond}.  Dimensional reductions of type (a) above can be
established by considering the density of states (DOS) as we will explain
below. Systems such as those discussed in Section \ref{exact_reduction}
and Kitaev's and Wen's models \cite{kitaev,wenmodel} illustrate
dimensional reductions of type (b) \cite{NO}. In such cases, similar to
\cite{bondprl,bond} a (unitary) transformation exactly maps the two
systems onto one another and preserves their spectrum. 

\section{Exact dimensional reduction from the density of states}
\label{N}

We will now illustrate how a technique based on the DOS
as applied to $n$-component vector theories (i.e., ``$O(n)$ theories'')
will enable us to establish dimensional reductions in various instances.
This includes some limits of $O(n)$ vector theories  such as that of
{\em large-$n$} or that of {\em weak coupling} or {\em high
temperatures}. In a Feynman diagrammatic framework, these relate to
situations in which we only consider  {\em tree level diagrams}.  As
emphasized earlier, although symmetries lead to bounds on
correlators such as those of Eqs. (\ref{strong}, \ref{g_n}), they
generally do not lead to an {\it exact} dimensional reduction per se.
That is, the thermodynamic functions in systems with a $\bf
d$-dimensional symmetry are, generally, not those of canonical $\bf
d$-dimensional systems but rather reflect the full dimensionality of
space-time.   There are, however, very general situations in which the 
dimensional reduction is guaranteed to be exact thermodynamically
(i.e., the thermodynamic functions of the $D$-dimensional systems are
identical to those of $d$-dimensional systems). 

We next demonstrate that in the large-$n$ limit of all vector theories,
the spatial dimensionality of the system can be changed at will if the total
DOS profile is kept the same.

\subsection{Preliminaries and statement of the DOS theorem}

We now focus on lattice field theory Hamiltonians of the form 
\begin{eqnarray}
H &=&  \frac{1}{2} \sum_{{\x},{\y}}  J({\x}-{\y})
{\bf{\phi}}({\x}) \cdot {\bf \phi}({\y})  + u \sum_{{\x}} ({\bf
{\phi}}^{2}({\x})-1)^{2},
\label{hk}
\end{eqnarray}
where  (i) the pertinent vector fields (${\bf{\phi}} = (\phi_1, \phi_2,
\cdots, \phi_n)$)  in Eq. (\ref{hk}) have $n \gg 1$ (large $n$)
components and (ii) the strength of the quartic coupling in Eq.
(\ref{hk}) is set by $u = 1/n$. In what follows, we denote the Fourier
transform of the translationally invariant $J({\x}- {\y})$ by $J({\bf
k})$. 

Given the general form of Eq. (\ref{hk}), we define the DOS via
\begin{eqnarray}
\label{dos}
g_{0}(\epsilon) \equiv  \int \frac{d^{D}k}{(2 \pi)^{D}} \
g_{0}(\epsilon, {\bf k}) ,
\end{eqnarray} 
where the {\it local} DOS is $g_{0}(\epsilon, {\bf
k})=\delta(\epsilon-J({\bf k}))$.   In Eq. (\ref{dos}), the $k$-space 
integration is performed over the first Brillouin zone for lattice
theories. In the regularized continuum field theoretic rendition of 
Eq. \eqref{hk}, the integration  is
performed over all real vectors ${\bf k}$.  As we will briefly discuss,
the DOS of Eq. (\ref{dos}) is that of the non-interacting
variant of Eq. (\ref{hk}), i.e.,  one with $u=0$. 

With these preliminaries, we are now ready to state our next theorem:
\vspace*{0.2cm}

\noindent
{\bf Theorem 2 (DOS):}
{\em The large-$n$ vector system of Eq. (\ref{hk}) in $D$ dimensions and
another large-$n$ vector system of the same form  in which the real
space interaction kernel  is replaced by another kernel that resides in
$D'$ dimensions,  $J({\x} - {\y}) \to \tilde{J}({\x'} - {\y'})$, will
have identical  free energies if their DOS are equal,
i.e., if} 
\begin{eqnarray}
\label{equivg}
\int d^D k ~\delta(\epsilon -J({\bf k})) = \int d^{D'}k'
~\delta(\epsilon-\tilde{J}({\bf k'})).
\end{eqnarray} 
\vspace*{0.2cm}

The proof of this assertion follows from rather standard results that we
show below. We first remark that this constitutes an extension of the
trivial variant of this statement for the non-interacting system of
$u=0$. In such a non-interacting system, Eq. (\ref{dos}) provides the
exact DOS for a single mode (and, by extension, to the
collection of decoupled modes that constitute  the non-interacting
system). Identical forms appear for other systems of decoupled particles
as in, e.g., the DOS for free electrons wherein the kernel
$J({\bf k})$ is replaced by a band dispersion $\epsilon ({\bf k})$. For
the lattice systems that we consider, each Fourier mode constitutes such
a free particle; the number of such particles (which we generally
generally denote by $N$) is equal, for the lattice systems that we focus
on, to $N_{s}  (\equiv |\Lambda|)$, the number of sites of the lattice.
The total DOS is given by 
\begin{eqnarray}
g(E) = \int d \epsilon_{1} \cdots \int d \epsilon_{N} ~
\delta(E-\epsilon_{1}  - \epsilon_{2} - \cdots - \epsilon_{N}).
\end{eqnarray}
Two systems are dual to each other and share the same spectrum if and
only if their densities of states $g(E)$  are identical  \cite{bondprl}.
Thus, in the non-interacting limit, when Eq. (\ref{equivg}) holds, the
two corresponding systems are trivially dual to one another.  The
special character of the large-$n$ vector theories renders this result
also precise {\it also when interactions  are introduced} ($u = 1/n$). 

The crux of our proof of the DOS theorem for large-$n$ vector theories,  and
more general possible extensions thereof  to other arenas, is that 
the partition function  depends solely on the DOS of the
system 
\begin{eqnarray}
{\cal Z} = {\cal Z} [g(E)].
\label{dos_inv}
\end{eqnarray}
Indeed, by construction, as ${\cal Z} = \tr_\Lambda (\exp(- \beta H))  = \int
d E \, g(E) \exp(-\beta E)$, the partition function is the Laplace
transform of the full exact DOS of the system $g(E)$.  As
such, any transformation that leaves the density of  states of the full
system $g(E)$ invariant automatically ensures that the  free energy is
the same. Two different systems of free particles (i.e., quadratic
systems) with an identical single particle DOS
$g_{0}(\epsilon)$ (that determines the full DOS of the
system) must, by Eq. (\ref{dos_inv}), display an identical density of
states $g(E)$ for the entire system. In such (free-particle type)
Gaussian systems, the partition functions can be further easily computed
in the presence of external sources. That leads to a
quadratic coupling between the  sources.  In interacting systems,
finding such a description with an effective density  of states $g_{\sf
eff}(\epsilon)$ is, generally, far more complicated. The dependence of
the full interacting system on a single density function  is reminiscent
of the density functional theory that has been of immense success in
solving quantum many-body systems. The Hohenberg-Kohn \cite{kohn}
theorems that encapsulate this approach illustrate how the system
properties may be effectively captured by a single density function.  A
physically appealing property of the dimensional reduction  that we
outline here is that, while preserving the DOS functional
(set by $g_{0}(\epsilon)$ in large-$n$ vector systems),  we may   map a {\it
local} system (one with interactions that decay with distance) in some
spatial dimension onto another {\it local} system residing in a  lower
spatial dimension.  As outlined above (and below), this preservation of
the spectrum, i.e., the DOS, while maintaining the local
character of the theory by performing such a map constitutes the key
physical underpinning of dualities \cite{bondprl}. 

We will next review some results concerning the large-$n$ vector
theories to explicitly illustrate how the result of the DOS theorem is
manifest. 

\subsection{Free energy of large-$n$ vector systems}

The DOS theorem follows nearly immediately from a diagrammatic calculation 
\cite{longdiag} as well as the well known equivalence
between large-$n$ vector theories to the spherical model \cite{Ma0}.  

 Specifically,
the free energy of (the to be reviewed) Eq. (\ref{free_D}) wherein 
$\mu$ is implicitly determined by Eqs. (\ref{self}, \ref{IM}) is 
manifestly dependent only on the DOS $g_{0}(\epsilon)$. 
This result similarly follows from diagrammatic considerations. 
Physically, as will elaborate later on in an example,  the
quintessential low-energy features of the theory are determined by the
soft modes.  

We review several known key features. In the large-$n$ limit, 
``hard spin'' models (those with $u \gg 1$ in Eq. (\ref{hk}) in which the fields 
have a fixed hard constraint on their norm, i.e.,  $|{\bf{\phi}}|=1$) tend to behave 
as their ``soft spin'' counterparts (with $u \sim 1/n$).   Furthermore, 
the large-$n$  rendition of Eq. (\ref{hk}) is, in many respects,
equivalent \cite{hes} to  that of the spherical model \cite{bk1952}.  In particular,
the  Helmholtz free energy density $F_{n}$ associated with the $O(n)$
system of Eq. (\ref{hk}) is, in the limit $n \to \infty$, identical to
the free energy density $F_{\sf spherical}$ of the spherical model
version of the $O(n)$ theory, i.e.,  $F_{n} = F_{\sf spherical}$.
On a finite size lattice, the $n$-component spherical model \cite{spheren}  is defined as a theory of
$O(n)$ fields with Hamiltonian 
\begin{eqnarray}
\label{hon}
H = \frac{1}{2} \sum_{{\x},{\y}}  J({\x}-{\y}) {\bf{\phi}}({\x}) \cdot
{\bf{\phi}}({\y}) =  \frac{1}{2N_{s}} \sum_{\bf k} J({\bf k})
|{\bf{\phi}}({\bf k})|^{2},
\end{eqnarray}
augmented by the {\it spherical constraint},
\begin{eqnarray}
N_s = \sum_{\x} {\bf{\phi}}^{2}({\x}).
\label{spherical_constraint}
\end{eqnarray}
The sum over wavevectors ${\bf k}$ is constrained to those in the first
Brillouin zone. In the thermodynamic limit, this sum may be replaced by
an integral. We may implement the constraint of Eq.
(\ref{spherical_constraint}) via a Lagrange multiplier $(\mu/2)$.  Thus,
in the large-$n$ limit, the quartic term of Eq. (\ref{hk}) may be
replaced by the spherical constraint of Eq. (\ref{spherical_constraint})
and the theory is defined by the Hamiltonian 
\begin{eqnarray}
H_{\sf spherical}= \frac{1}{2} \sum_{{\x},{\y}}   J({\x}-{\y}) 
{\bf{\phi}}({\x}) \cdot {\bf{\phi}}({\y}) +  \frac{\mu}{2}  \sum_{{\x}}
{\bf{\phi}}^{2}({\x}) .
\label{sphere1}
\end{eqnarray}

In order to compute averages, we augment $H_{\sf spherical}$ by local  
fields ${{h}}({\x})$ that couple linearly to the vector  field
${\bf{\phi}}(\x)$ and take functional derivatives of the resulting
generating functional ${\cal Z}$ relative to these fields.  Thus, in the
spherical model, or lowest (zeroth) order $1/n$ computation, a Gaussian
integration over the fields ${\bf{\phi}}$ leads to the corresponding
canonical generating functional
\begin{eqnarray}
{\cal Z}_{\sf spherical} &=&  \int  D  {\bf{\phi}} \exp(-\beta [  H_{\sf
spherical}-  \hspace*{-0.1cm}\sum_{{\x}}  h({\x}) \cdot \phi({\x})] )
\nonumber \\ 
&=&   \prod_{{\bf k}} \sqrt{\frac{2 \pi}{\beta (J({\bf k}) + \mu)}} \
\exp[-\frac{\beta}{2} \sum_{{\bf k}} \frac{h({\bf k}) \cdot 
h(-{\bf{k}})}{J({\bf k})+ \mu}],
\label{Zp}
\end{eqnarray}
where $D {\bf{\phi}}$ is a shorthand for the lattice measure 
$\prod_{\x}^{N_{s}} d \phi({\x})$.  By differentiation relative to the
fields ${h}({\bf k})$  (or, alternatively, by the use of the
equipartition theorem), it is seen that
\begin{eqnarray}
\langle |{\bf{\phi}}({\bf k})|^{2} \rangle =  \frac{1}{\beta (J({\bf k})
+ \mu)}.
\label{eqp}
\end{eqnarray}
Inserting Eq. (\ref{eqp}) into Eq. (\ref{spherical_constraint}) and
accounting for a possible condensate (detailed below), we arrive at the
self-consistency equation
\begin{eqnarray}
\beta &=& \int \frac{d^{D}k}{(2 \pi)^{D}} \frac{1}{J({\bf k}) + \mu} +
\beta I_{M}. 
\label{self}
\end{eqnarray}
The last term represents, at temperatures below  $T_{c}$ (if a critical
temperature exists),  the condensate of modes ${\bf q}$ that lie on the
manifold of minimizing modes $M$.  That is,  $J({\bf q} \in M) =
\min_{{\bf k}} \{ J({\bf k}) \}$ and in Eq. (\ref{self}),
\begin{eqnarray}
\label{IM}
I_M = N_s^{-1} \sum_{{\bf q} \in M}\langle  |\phi({\bf q})|^{2} \rangle.
\end{eqnarray}  
As is evident from Eq. (\ref{Zp}), a permutation $P$ of the
wavevectors  $\bf k$ (i.e., a transformation $J({\bf k}) \to J(P{\bf
k}) = J({\bf k'})$) will leave  the product over all $\bf k$
invariant.   The generating functional depends only on the density of
states $g(\epsilon)$.  Then, it is clear that the zero field  (${h}=0$)
free energy  is given by  
\begin{eqnarray}
F_{\sf spherical}(h=0) = -\beta^{-1} \ln {\cal Z}_{\sf spherical}({h}=0)
=  \frac{1}{2 \beta} \int d \epsilon~ g_{0}(\epsilon) \ln[\beta(\epsilon +
\mu)],
\label{free_D}
\end{eqnarray}
which clearly demonstrates the DOS theorem  of Eq. (\ref{equivg}).

These results also follow from the direct diagrammatic expansion   of
the theory of Eq. (\ref{hk}) to lowest order  in $(1/n)$ (i.e., to order
$O((1/n)^{0})$) \cite{longdiag,Ma}.

\subsection{Examples of dimensional reduction in large-$n$ theories}

We should stress that the dimensional reduction that arises in the
large-$n$ theory is general and not limited to these examples. Below, we
will review facts pertaining to  the DOS in canonical
systems with a quadratic dispersion about the minimum of the energy in
$k$-space and employ these in two examples of dimensional reductions
that highlight quintessential aspects of the DOS theorem. Insofar as the
simple theory of  Eq. (\ref{compass}), an explicit calculation in the
large-$n$  limit illustrates that this system is identical to a
one-dimensional system in the same limit. We now provide two simple (yet
slightly less trivial) examples.

{\bf Example 1:}   Consider the case of nearest-neighbor interactions on
a hypercubic lattice in $D$ dimensions in which the lattice constant $a$
is set to one. In such a case, a ferromagnetic lattice large $n$ model
is given by Eqs. (\ref{hon}, \ref{sphere1}) with $J({\x}-{\y}) = - J$ if
sites ${\x}$ and ${\y}$ are nearest neighbors and is zero otherwise.  
In Fourier space, the corresponding kernel is given by
\begin{eqnarray}
J({\bf k}) =  2J \sum_{\ell=1}^{D} (1- \cos k_{\ell}) \equiv J
\Delta({\bf k}).
\label{jk}
\end{eqnarray}
As seen, here, $\Delta({\bf k})$ denotes the lattice Laplacian in $D$
dimensions. In the low energy or the small $\bf k$ (long-wavelength)
limit $\Delta({\bf k})  \to { k}^{2}$.  The number of points in
$k$-space that have a given value of $J({\bf k})=\epsilon$ is set by the
surface area of a sphere whose radius is given by $(\epsilon/J)^{1/2}$.
Thus, the DOS scales as  $g_{0}(\epsilon) \sim
\epsilon^{(D-1)/2}$. In  Eq. (\ref{jk}), the quadratic minimum (for
$J>0$) occurs at a single point in $k$-space, ${\bf k}={0}$.

Let us start by considering the case in which the kernel $J({\bf k})$ of
Eq. (\ref{hk}) has its minimum on a manifold $M$ of dimension $d_{\min}$ and
that $J({\bf k})$ exhibits a quadratic dispersion about this minimum. In
such cases, the DOS is that of a $D'= (D-d_{\min}$)
dimensional system.  That is, to leading order 
\begin{eqnarray}
g_{0}(\epsilon) \sim \epsilon^{(D'-1)/2}.
\label{ged}
\end{eqnarray}
The dependence of the generating functional solely  on the density of
states $g_0(\epsilon)$  allows all translationally invariant systems
with a particular momentum space dispersion $\epsilon_{k}$ to  be mapped
onto a one-dimensional system.  Towards this end, we define a kernel
$J_{\sf eff}(k)$  of an effectively equivalent one-dimensional system
(i.e., that with a scalar $k$) by
\begin{equation}
\int \frac{d^{D}k}{(2\pi)^D}  \ \delta(\epsilon-J({\bf k})) = \left  
|\frac{dJ_{\sf eff}}{dk} \right |_{J_{\sf eff}(k)=\epsilon}^{-1}.
\end{equation}
The above relation ensures that the DOS and consequently
the generating functional are preserved.   That is, the generating
functional of the $D$-dimensional system with the kernel $J({\bf{k}})$
is identical to that of a one-dimensional system with the kernel $J_{\sf
eff}(k)$. 

As an example, we now consider the nearest-neighbor system of Eq. (\ref{jk}) 
in two dimensions ($D=2$). We will now map this system onto 
a $d=1$ dimensional  system. Towards this end, we write
\begin{eqnarray}
 \left  |\frac{dJ_{\sf eff}}{dk} \right |^{-1}_{J_{\sf eff}(k)=\epsilon}
 &=& \int_{-\pi}^{\pi} \frac{dk_{1}}{2 \pi}  \int_{-\pi}^{\pi}
 \frac{dk_{2}}{2 \pi} ~ \delta(\epsilon- 2J(\cos k_{1} + \cos
 k_{2}))\nonumber \\ 
 &=& \frac{1}{4 \pi^{2} J} \int_{0}^{1} \frac{dv}{\sqrt{1-v^{2}}
\sqrt{1-(\tilde{\epsilon}-v)^{2}}} = g_{0}(\epsilon),
 \label{triv_long1}
\end{eqnarray}
with $\tilde{\epsilon} = \epsilon/(2J)$. The second equality in Eq.
(\ref{triv_long1}) trivially follows from, e.g., the substitution $v=
\cos k_{1}$ (after $w= \cos k_{2}$ is integrated over and consequently
set by the delta function).  Thus,  
\begin{eqnarray}
k(J_{\sf eff})=  \int_{0}^{J_{\sf eff}}  d \epsilon ~ g_{0}(\epsilon) .
  \label{elliptic}
\end{eqnarray}
Equation (\ref{elliptic}) may be inverted and Fourier transformed to
find the effective one-dimensional  real space kernel $J_{\sf
eff}(x-y)$. By inserting this back into Eq. (\ref{hk}), we complete the
mapping of the two-dimensional  nearest-neighbor system onto a
one-dimensional  system. In a similar fashion, within the spherical  (or
equivalently the $O(n \rightarrow \infty)$) limit, all high-dimensional
problems may be mapped onto a translationally invariant one-dimensional 
problem. It is important to re-iterate, in light of the example provided
above, that the lower (one-) dimensional model albeit having longer
range interactions than its higher dimensional counterpart (in fact,
scaling as $V(r) \sim r^{-2}$ for large $r = |x-y|$), the resulting
theory is nevertheless {\it local} and physical. 

{\bf Example 2:} As {\em another more natural example where dimensional
reduction occurs},  we consider next a situation wherein $J({\bf k})$
attains its global minimum on a spherical  shell $|{\bf k}| =q>0$. 
When $J(\bf{k})$ attains its minimum on a spherical
shell, then in an expansion of $\epsilon_{\bf k}$ about any such point, 
there is only one direction in ${k}$-space (the radial
direction) along which the dispersion may be quadratic and $(D-1)$
tangential directions with {\it soft modes}. The DOS is
determined solely by the massive radial direction, and  this system
behaves as a one-dimensional one. The correlation functions  are given
by derivatives of the generating functional relative  to ${h}$ in which
the limit ${h} \to {0}$ is taken. If the low energy dispersion is given
by Eq. (\ref{ged}), then the free energy (in which only states in the
vicinity of the minimizing manifold are considered as is appropriate at
low temperatures) will have the canonical form of a $ D' = D-d_{\min}$
dimensional system. That is so as in the coordinate system locally
tangential to the  minimizing manifold in $k$-space, there are $D$ 
directions on which the dispersion $J({\bf k})$ has a quadratic
dependence and $d_{\min}$ directions on which  it has no dependence.
Thus, in the large-$n$ limit, not only are some correlation functions of
a form characteristic to that of a (lower) $D'$-dimensional system as we will
show but also the thermodynamic functions themselves are also
characteristic of a low-dimensional system. Using Wick's theorem for the
quadratic large-$n$ theory, the correlation functions between $2m$
fields are given by
\begin{eqnarray}
 \langle \phi({\bf k}_{1}) \phi({\bf -k}_{1}) \cdots \phi({\bf k}_{m})
\phi({\bf -k}_{m}) \rangle = \prod_{i=1}^{m}  \frac{1}{ \beta(J({\bf
k}_{{i}}) + \mu)}.
\end{eqnarray}

Cast in terms of symmetries, the operation that enables us to reduce the
$D$-dimensional problem onto a $D'$-dimensional one is the {\em presence
of a non-trivial permutational $d_{\min}$-dimensional symmetry  of the
spectra}. That is, any permutation $P$ (whose minimal non-empty set
spans manifolds $M$ of dimension $d_{\min}$) that  maps wavevectors
${\bf{k}}$ with the same energy ($J(\bf{k})$) onto each other is a
symmetry. The full $d_{\min}$-dimensional symmetry enables us to
reorganize the labeling of states in $k$-space such that the resultant
problem after a permutation corresponds to a $D'$-dimensional one (or
more precisely, at low temperatures $|M|$ copies of a $D'$-dimensional
system where $|M|$ is the number of minimizing  modes). We briefly elaborate
on this. We can choose permutations contort the general
$d_{\min}$-dimensional equipotentials of constant $J$ onto a set of flat
$d_{\min}$-dimensional hyperplanes. These represent a stack of decoupled
$D'$-dimensional  systems (the stacking occurs along $d_{\min}$
directions along which there are no interactions: there is no mode
dispersion).  

\subsection{Weak coupling or high temperature}

We now discuss an extension of the DOS theorem of Eq. (\ref{equivg}) to
another class of systems.  As illustrated in   \cite{hight},  in the
high temperature (or weak coupling) limit of any $O(n)$ theory (with an
arbitrary finite number of components $n$) of the  form of Eq.
(\ref{hk}),  the free energy is given by 
\begin{eqnarray}
\label{free}
F = \frac{1}{2\beta} \sum_{{\bf{k}}} \ln[\frac{\beta}{2 J({\bf k})} +1].
\end{eqnarray}
Thus, similar to the case of the large-$n$ vector theories, all that
matters in Eq. (\ref{free}) is the DOS $g(E)$. In
particular, any system can have its dimensionality reduced.  There are
numerous systems that, at the high-temperature (small $\beta$) limit
have an identical free energy. A subset of these systems become lower
dimensional both (i) in the limit of small $\beta$ (for finite $n$) and
(ii) the limit of large $n$ for finite $\beta$.  For instance, a
one-dimensional Ising  chain with a momentum space interaction kernel
given by Eq. (\ref{elliptic}) with an Ising version of Eq. (\ref{hon})
is identical, in the limit of high temperatures, to the two-dimensional
nearest-neighbor Ising model. 

\subsection{Dimensional reductions and density of states functional
theories}

Beyond the particular confines of the large-$n$ vector theories and high
temperature/weak coupling limits, the likes of the DOS theorem might hold
elsewhere. We now make general remarks about the general premise and
viable extensions  of the DOS theorem. We first couch our specific
considerations thus far  within a simple standard framework. As
highlighted in Eq. (\ref{dos_inv}), in these theories, a single (DOS)
functional determines the system properties as the DOS
determines the generating (or partition) function ${\cal Z}$. As
it so happens in the $O(n)$ theories that we looked at the free energies
at hand were a DOS functional  set by the
interaction kernel in $k$-space.  The DOS theorem would hold without any
change for more general functionals. We
remark that density functional theories with correlations set by the
local DOS in real space representations (au lieu of the
$k$-space dependence that investigated in the large-$n$ theories)  have
been advanced in solid state systems \cite{soler}.   Just as
in many-body systems, it well may be hard to determine the DOS
functional apart from easier cases such as that of large $n$. 

We speculate that a dependence on
the metric $g_{\mu \nu}(\x)$ in gravity theories might play an analogous role to the
dependence on the interaction kernels that we investigated above for
large-$n$ vector theories, and that effective {\em metric functional
theories} might exist by an extension of standard methods of density
functional theories \cite{kohn}. In such instances, computable
quantities would be set  by functionals of the metric. 
We would like to emphasize, though, that
an equivalent of the Hohenberg-Kohn theorems should be proved first. 
Had this proof be non-existent then there is no possibility of an exact duality 
(i.e., an isomorphism) between quantum many-body or non-relativistic field theories
and semiclassical gravity theories. It could still happen that a duality 
{\it emerges} in some appropriate limit. 

We now step back to highlight the specifics of the core features of the
derivation of the DOS theorem and the applications that we provided. Our
treatment was centered on  the Gaussian type character of large-$n$
theories.   In Eq. (\ref{Zp}), we introduced sources. The functional
derivatives relative to the  source fields $h$ yield the correlation
functions. Written in a more general form, Gaussian integration leads to
a free (i.e., quadratic order) form in the source fields which  are
coupled via the correlation function $G_0^{(2)}$. The standard large-$n$
results  that we invoked and expanded on may be more generally  linked to a
standard perturbative framework. Within a general perturbative framework
the kernel $[J({\x} - {\y}) + \mu]$ of Eq. (\ref{sphere1}) coupling the
fields $\phi({\x})$ and $\phi({\y})$ may, more generally, be replaced by
a zeroth order (i.e., quadratic) two-point kernel 
$[(G_{0}^{(2)})^{-1}]$. In such a case,  the generating functional is
\begin{eqnarray}
{\cal Z}_0(\beta,h)= {\cal Z}_{0}(\beta,0)  \exp \left [-
\frac{1}{2}\int d^D x \, d^D y ~  h ({\x})  (\beta G_{0}^{(2)}
({\x},{\y})) h ({\y})\right ].
\label{Gauss2}
\end{eqnarray}

As is  well known, for a general  Hamiltonian (or action)
density, which is the sum of a quadratic and an interacting part ${\cal
H}_{\sf int}$,  the corresponding generating functional becomes
\begin{eqnarray}
\label{zhg}
{\cal Z}(\beta,h)=\exp \left [- \int d^D x ~\beta \ {\cal H}_{\sf int}
(- \delta/\delta (\beta h({\x})))\right ] {\cal Z}_0(\beta,h).
\end{eqnarray}
Thus, whenever dimensional reduction is exact on the quadratic level, it
may, in certain cases (e.g., that of Example 2 above) enable effective
dimensional reductions when a perturbative expansion about the quadratic
theory is valid. 
 
There is, generally, no expression for the lowest-order partition 
function in the presence of sources, e.g., Eq. (\ref{Zp}), in terms of 
the single particle DOS. In the presence of sources, the 
partition function depends on the full dispersion $J({\bf k})$ in 
${k}$-space and not solely on the single particle DOS 
$g_{0}(\epsilon)$ (nor the full DOS of the entire system 
$g(E)$ in the absence of sources). This feature underscores the larger 
information content that may be generally present in high-dimensional 
${k}$-space when sources are present, vis a vis that embodied in the 
system dependence only on the single energy coordinate $\epsilon$ (or
$E$).   We  employed this redundancy in the higher-dimensional
${k}$-space  in  order to freely move from a system in $D$ dimensions to
a system in a  lower dimension when sources were not present (Example
1). In  situations in which two dispersions $J({\bf k})$ and
$\tilde{J}({\bf  k'})$ in different dimensions (say $D$ and $D'$, with
$D>D'$)    exhibit identical behavior in the vicinity of their minima
then the   partition functions in the presence of sources and (the 
corresponding  correlation functions) may be identical.  This path was
realized in  Example 2 above. In such a situation, the high-dimensional
function  $J({\bf k})$ is constant, i.e.,  does not disperse, for
variations in ${\bf  k}$  along $D-D'=d_{\sf min}$ directions,  while
along $D'$ components of ${\bf k}$ the function $J({\bf k})$  is
identical to $\tilde{J}({\bf k'})$. As  $J({\bf k})$ captures, to lowest
order, the correlation function   $G_{0}^{(2)}$, two such systems
exhibit identical correlations.  Moreover, the identity of the 
dispersions $J({\bf k})$  and  $\tilde{J}({\bf k'})$ ensures that the
corresponding DOS are  the same and, thus,  the free
energies are identical.
 
We now briefly remark and speculate on possible connections between the
above results and those suggested by AdS-CFT correspondences  
\cite{maldacena}. The relation of Eq. (\ref{equivg}) enables a
construction of candidate gravity duals for $O(n)$ theories. 
Recently   \cite{sung}, an attempt has been made to find the gravity
dual  of an $O(n)$ vector theory using a renormalization group scheme. 
It was found that an $O(n)$ vector theory corresponds, via an AdS-CFT
type correspondence,  to gravity in the Arnowitt-Deser-Misner (ADM)
representation   \cite{adm} with two sets of conjugate fields. Perusing
the results of  Ref.   \cite{sung}, we note that in the large-$n$ limit,
the dual ADM theory is quadratic (as it must be) just as the original
large $O(n)$ field theory to which it is dual.  With the results of that
we derived in the current section,  it may be straightforward to check
whether the relation of Eq. (\ref{equivg}) is satisfied in this case.
This is so as our relation of Eq. (\ref{equivg}) enables the appearance
of dual equivalent theories of many sorts, not only dual gravity
theories to standard CFTs, that exist in one higher dimension.  

The {\it
ground state entropy of vector theories in the large-$n$ limit  has a
holographic character}, i.e.,  it scales as the surface area of the
minimizing manifold $M$ in $k$-space
\begin{eqnarray}
{\sf S} \sim L^{d_{\min}}.
\label{hologram}
\end{eqnarray}
As in our discussion of Example 2, in Eq. (\ref{hologram}),  $d_{\min}$
is  the dimension of the surface $M$ of minimizing modes in $k$-space
\cite{zn}.  The proof of this assertion is trivial. Any state 
$\phi({\bf x}) = \sum_{{\bf{q}} \in M} \phi({\bf{q}})  \exp [i {\bf{q}}
\cdot {\bf{x}}]$ (with $\phi(-{\bf q}) =  \phi^*({\bf q})$ to ensure
that $\phi({\bf x})$ is real) whose sole non-vanishing Fourier modes are
those for which $J({\bf k})$  attains its minimum is a ground state. The
manifold $M$ has dimension $d_{\min}$ and thus there are
$O(L^{d_{\min}})$ minimizing modes ${\bf{q}}$ and corresponding Fourier amplitudes
$\phi({\bf{q}})$. These amplitudes have to adhere to the single global
normalization constraint  of Eq. (\ref{spherical_constraint}). This
leaves a total of $O(L^{d_{\min}})$ free amplitudes  $\{\phi({\bf q})\}$
and to the sub-extensive holographic-like entropy of Eq.
(\ref{hologram}). For finite-$n$ systems, not only the density of
eigenmodes in ${k}$-space is important and the above correspondence
breaks down. 

\section{Exact dimensional reduction from bond algebras}
\label{exact_reduction}

In this section we will rely on a new bond-algebraic theory of
dualities \cite{bondprl,bond} to study exact dualities that connect 
models of dimension \(D\) to models of dimension \({\bf d}<D\).   These
dualities afford especially transparent examples of exact  dimensional
reduction.  Dualities have been the subject of intense research efforts
for some years now since the AdS-CFT correspondence is conjectured  to
be a particular example of a large
class of dualities that effect a change of dimension \cite{review, nastase}. 
The subjects of TQO and fault-tolerant quantum computation are also rich
in models in \(D\geq2\) dimensions that are dual to \({\bf
d}=1\)-dimensional models. This is not  surprising if one notices that
these \(D\)-dimensional models typically display {\bf d}\(=1\)-GLSs 
\cite{NO} and,  as we have seen already, {\bf d}-GLSs are closely linked
to effective dimensional reduction. However, the precise connection
between {\bf d}-GLSs and {\it exact} dimensional  reduction requires
further discussion. We would like to understand  whether 1) models that
display {\bf d}-GLSs are always exactly dual to  lower-dimensional
models, and conversely, whether 2) \(D\)-dimensional  models dual to
\({\bf d}<D\)-dimensional ones must display {\bf d}-GLSs. That {\bf
d}-GLSs do not always imply exact dimensional reduction  will be shown 
by  a counter-example in the next section. As for 2), we do not have a
full understanding  yet. 

There is a different aspect to the connection between effective and
exact dimensional reduction that also calls for clarification.
The bounds derived in Eq. \eqref{g_n} are not saturated in general,
meaning  that the upper and lower bounds $G^{(n)}_{{\sf upper}; {\bf
d}}$ and $G^{(n)}_{{\sf lower}; {\bf d}}$  may be different. They
nevertheless afford a sharp characterization of  effective 
dimensional reduction, and from 
this perspective, {\it exact} dimensional reduction may be understood as
the special case in which inequalities of the  form of Eq. \eqref{g_n}
become equalities. It is natural to conjecture that this happens if and
only if there is an exact duality that maps the \(D\)-dimensional model
to an explicitly lower dimensional model. The reason is that a duality
amounts to a unitary equivalence \cite{bondprl}. It follows that the
correlators of several appropriately defined quantities in a model of
dimension \(D>{\bf d}\) dual to a  model of  lower dimension \(\bf d\)
must show {\it exact} \(\bf d\)-dimensional behavior, because they can
be mapped to and represented by corresponding duals in the low ($\bf d$)
dimensional system. This is a fairly simple proof of the ``if" part of
the conjecture. At present, we do not know  beyond some specific
examples whether the ``only if" part holds as well.

Regardless of their connection to effective dimensional reduction, exact
dualities that change dimension will be studied here also for their own
sake. We will see that they represent a delicate phenomenon that results
from a very peculiar, hard to characterize structure of interactions
({\it bond algebra} in the language of \cite{bondprl,bond}),  and that
dual models of different dimension can only be modified in limited ways
without destroying the duality that connects them. We think that these
patterns visible in concrete examples should be kept in mind when
gauging the merit of conjectured dualities that change dimension, and
most importantly, when trying to extend these conjectures in non-trivial
ways.  

We will focus both for simplicity and for their own merit on lattice
models of TQO. If a system displays TQO, its state cannot be solely
characterized  by local measurements. Rather measurements of
topological quantities are needed too   \cite{wenbook,kitaev}.   
Systems characterized by  topological (non-local) order are, in
principle, robust to local perturbations and thus  able to store quantum
information for long periods of time. Such systems would be ideal
candidates for  realistic {\it quantum memories} for quantum information
processing  and computation. Unfortunately, many early models harboring
TQO were shown \cite{NO} to be dual to one-dimensional Ising chains with
short-range  interactions, a result that suggests that the
autocorrelation times in these particular models are system-size
independent. Moreover, at equilibrium all initial information becomes
random and is lost at {\it any} non-zero temperature,  a phenomenon
dubbed {\it thermal fragility}  \cite{NO}.  Here we will show that this
property  is also shared by several  new models that have been devised
and investigated since the early days of TQO. They are:
\begin{enumerate}
\item{ a $D=2$ quantum model on a honeycomb lattice 
generalizing the one introduced in Ref. \cite{ly},}
\item{$D=2$ topological color codes \cite{color}, and} 
\item{a $D=3$ model on an fcc lattice introduced in Ref.
\cite{chamonmodel} and further investigated in Ref. \cite{sergey}.}
\end{enumerate}
These models are all dual to one-dimensional Ising chains or ladders
with appropriate boundary conditions, and appropriately distributed
transverse fields.  They display exact dimensional reduction, and, as a
consequence,  memory times that are system-size independent at all
temperatures. 

In closing, let us notice that  the techniques and  ideas to be
discussed next are perfectly suitable \cite{bondprl} for studying models
of  lattice Hamiltonian quantum field theory \cite{susskind}. This
observation may become specially relevant if and when lattice renditions
of the AdS-CFT and related conjectures become available for direct
numerical testing.

\subsection{Extended toric code on a honeycomb lattice}
\label{etch}
  
\begin{figure}[htb]
\centering
\includegraphics[angle=0, width=.5\columnwidth]{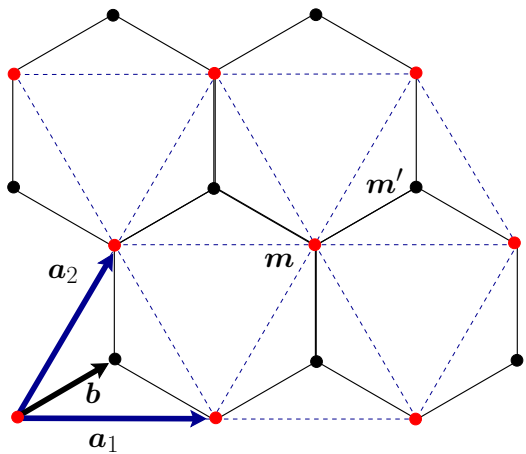}
\caption{A honeycomb lattice described as a triangular Bravais lattice
with a basis.}\label{hex_lat_basis}
\end{figure}

The model that we are going to study in this section is a variation of a
model introduced in Ref. \cite{ly}. It features \(S=1/2\) spins  on each
site  of a honeycomb lattice,  and shows exact dimensional reduction
despite having non-commuting operators in its Hamiltonian.

The honeycomb lattice can be described as a triangular  Bravais
lattice,  generated by the primitive vectors \(\a=\i\) and 
\(\aa=(\i+\sqrt{3}\j)/2\)  
(${\i}$ and ${\j}$ are the standard unit  vectors of a square lattice),  
together with a basis described by the
two vectors \(\bm{0}\) and \(\bm{b}=(\i+(\j/\sqrt{3}))/2\), see Fig.
\ref{hex_lat_basis}. The sites of the lattice are spanned by the sets of
points \(\m=m^1\a+m^2\aa\), and the set \(\bm{m'}=\m+\bm{b}\), with
\(m^1,m^2\in \mathbb{Z}\). Imposing periodic boundary conditions amounts
to  identifying two sites \(\m_1,\m_2\) to be the same if their
coordinates \(m^\mu_1,m^\mu_2\) are equal up to a multiple of the period
\(L_\mu,\ \mu=1,2\); and we identify two sites \(\m_1',\m_2'\) if
\(\m_1'-\bm{b},\m_2'-\bm{b}\) satisfy the same condition. The result is
a honeycombl lattice \(\Lambda^P_{\sf hex}\) with the topology of a 
two-dimensional torus. Since we can associate each elementary hexagon 
to a site \(\m\) in the triangular sublattice, we can use \(\m\) as  an
index to label the two plaquette operators
\begin{eqnarray}
&&W_{(1,\m)}=\sigma^z_\m\sigma^x_{\bm{m'}}\sigma^y_{\m+\aa}
\sigma^z_{\bm{m'}+\aa-\a}\sigma^x_{\m+\aa-\a}\sigma^y_{\bm{m'}-\a}\ ,
\label{w1} \nonumber\\
&&W_{(2,\m)}=\sigma^z_\m\sigma^y_{\bm{m'}}\sigma^x_{\m+\aa}
\sigma^z_{\bm{m'}+\aa-\a}\sigma^y_{\m+\aa-\a}\sigma^x_{\bm{m'}-\a}\ .
\label{w2}
\end{eqnarray}
where \(\sigma^\alpha_\m\) with $\alpha=x,y,z$ are Pauli matrices.
Notice that {\it any} two of these plaquette operators commute,
$[W_{(\mu,\m)}, W_{(\nu,\bm{n})}]=0$.

\begin{figure}[htb]
\centering
\includegraphics[angle=0,width=.40\columnwidth]{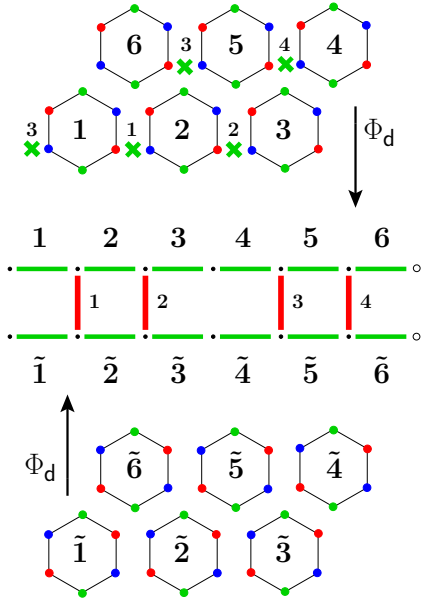}
\caption{Honeycomb lattice with \(L_1=3,\ L_2=2\), and periodic boundary
conditions (twelve distinct sites). The top panel shows the plaquette
bonds \(W_{(1,\m)}\) introduced in Eq. \eqref{w1}, numbered from  one to
six, and the four transverse fields \(\sigma^z\) as green crosses (the
spin labelled as three is not part of the Hamiltonian, but it appears in
one of its ${\bf d}$-GLSs). The bottom panel shows the plaquette  bonds 
\(W_{(2,\m)}\). The green (blue, red) dots stand for spins \(\sigma^z\)
(\(\sigma^x,\ \sigma^y\)).   The middle panel represents the dual,
one-dimensional ladder model defined in Eq. \eqref{dual_ely}. The
duality mapping  \(\Phi_\d\) defined in Eq. \eqref{d_iso_ly} is
indicated by corresponding labels in the illustrations of both models.
For example, the plaquette bond in the bottom panel labelled  as
\(\tilde{1}\) maps to the link \(\sigma^z_{2,1}\sigma^z_{2,2}\)
indicated by a green horizontal  bar, and labelled by \(\tilde{1}\) as
well. The dual model satisfies periodic boundary conditions. 
}
\label{ely_duality}
\end{figure}

The honeycomb extended toric code (HETC) model that we will
study is specified  by the Hamiltonian 
\begin{equation}\label{ely}
H_{\sf HETC}=-\sum_{\mu=1,2}\sum_{\m\in \Lambda^P_{\sf hex}}\ J_\mu
\,W_{(\mu,\m)}
-h\sum_{m^1=1}^{L_1-1}\sum_{m^2=1}^{L_2}\sigma^z_{\bm{m}},
\end{equation} 
illustrated in Fig. \ref{ely_duality} for a lattice with periods
\(L_1=3,\ L_2=2\). The Hamiltonian acts on a Hilbert space of dimension 
\((2^{N_s})^2=2^{2N_s}\), with \(N_s=L_1L_2\). If \(h=0\), \(H_{\sf
HETC}\) reduces to the model introduced in Ref. \cite{ly}. Notice that
the magnetic field \(h\) is staggered, meaning that it is zero on all 
sites \(\m'\), and that it is  mostly homogeneous, except for the
(one-dimensional) circle of sites  \(L_1\a+m^2\aa,\ m^2=1,\cdots, L_2\),
where it is also zero. The Hamiltonian  \(H_{\sf HETC}\) is self-dual
under the exchange  \(J_1\leftrightarrow J_2\), but we will not show
this directly, as  it will become clear after the duality to a \({\bf
d}=1\) model.

We will use the new approach to dualities developed in Refs.
\cite{bondprl} to look for dual representations of \(H_{\sf HETC}\).
Consider the elementary operators \(W_{(\mu,\m)}\) and \(\sigma^z_\m\),
and only those operators. It is convenient to have a uniform
denomination for them, and so we agree to call them generically {\it
bonds} \cite{bond}. The bonds of a Hamiltonian (like \(H_{\sf HETC}\))
generate a  model-specific bond algebra of interactions that has
typically a very different structure  as compared to  the algebra of
elementary degrees of freedom (spins $S=1/2$ in the case at hand). So it
was proposed in \cite{bondprl} to look for dualities as
structure-preserving mappings (homomorphisms, or isomorphisms in the
absence of gauge symmetries) of bond algebras local in the bonds. To
find a dual representation of  \(H_{\sf HETC}\) we need to characterize
its bond algebra, and find a model \(H_{\sf HETC}^D\) that looks
different but has in fact a bond algebra isomorphic to that of \(H_{\sf
HETC}\).

Since each bond \(\sigma^z\) anticommutes with four plaquettes,  
\(H_{\sf HETC}\)'s bond algebra is non-commutative.  The bonds
\(W_{(\mu,\m)}\) commute with each other, and satisfy the constraints
\begin{equation}
\prod_\m\ W_{(\mu,\m)}=\mathds{1},\ \ \ \ \mu=1,2.
\end{equation}
Keeping these relations and constraints in mind, one can check that the
bonds for a one-dimensional ladder of spins with periodic boundary
conditions satisfy  identical relations
and constraints, as illustrated in Fig. \ref{ely_duality}.  It follows
that  \(H_{\sf HETC}\) is dual to 
\begin{eqnarray}\label{dual_ely}
H_{\sf HETC}^D=-\sum_{\mu=1,2}\sum_{m=1}^{N_s}\ J_\mu
\sigma^z_{\mu,m}\sigma^z_{\mu,m+1}
-h\sum_{s=0}^{L_2-2}\sum_{i=2}^{L_1-1}
\sigma^x_{1,i+s(L_1-1)}\sigma^x_{2,i+s(L_1-1)}.
\end{eqnarray}
This establishes the one-dimensional character of \(H_{\sf HETC}\). 
The  self-dualities of that Hamiltonian follow from those of the Ising ladder.

\begin{figure}[htb]
\centering
\includegraphics[angle=0,width=.6\columnwidth]{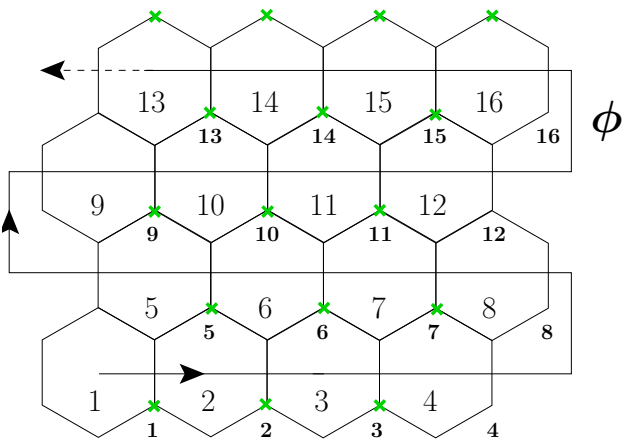}
\caption{Numbering function \(\phi\) used to describe the duality
between \(H_{\sf HETC}\) and a one-dimensional model, illustrated for  a
lattice with \(L_1=4\) and \(L_2> 4\). \(\phi\) numbers the hexagons, or
equivalently, the sites of the underlying triangular Bravais lattice
(large  numbers). Since there is one point of the basis per hexagon,
\(\phi\) numbers these as well (small numbers). The green crosses
represent the  transverse-field bonds \(\sigma^z\) present in \(H_{\sf
HETC}\). Periodic boundary conditions are considered.}
\label{numbering_hex}
\end{figure}

It will be useful in the following to have an explicit description 
of the duality mapping connecting the two models of
Eqs. (\ref{ely}, \ref{dual_ely}). First, we number the sites of the 
honeycomb lattice as in Fig. \ref{numbering_hex}. The duality mapping then 
reads
\begin{eqnarray}\label{d_iso_ly}
W_{(\mu,\m)}&\dual&\sigma^z_{\mu,\phi(\m)}\sigma^z_{\mu,\phi(\m)+1},\\
\sigma^z_{\bm{m'}}&\dual&
\sigma^x_{1,\phi(\bm{m'})+1}\sigma^x_{2,\phi(\bm{m'})+1}, \nonumber
\end{eqnarray}
so that \(\Phi_\d(H_{\sf HETC})= H_{\sf HETC}^D\), see Fig. \ref{ely_duality}. 
The theory of operator algebras guarantees that
isomorphisms like \(\Phi_\d\) of Eq. \eqref{d_iso_ly} must
be equivalent to a unitary transformation \cite{bondprl}. 
It follows that there is a unitary transformation \(\mathcal{U}_\d\) such
that \(H_{\sf HETC}^D=\Phi_\d(H_{\sf HETC})=\mathcal{U}_\d
H_{\sf HETC}\mathcal{U}_\d^\dagger\).

The Hamiltonian \(H_{\sf HETC}\) displays ${\bf d}=1$-GLSs that map
into  symmetries of the dual ladder model. Consider for concreteness the
model  illustrated in Fig. \ref{ely_duality}. It is clear that the
operator
\begin{equation}
\sigma^z_1\sigma^z_2\sigma^z_3
\end{equation}
(using the numbering of spins described in Fig. \ref{ely_duality})
commutes with \(H_{\sf HETC}\). To determine the dual symmetry
operator   \(\Phi_\d(\sigma^z_1\sigma^z_2\sigma^z_3)\) we need to
compute \(\Phi_\d(\sigma^z_3)\), which can be done according to the
techniques of Ref.   \cite{bondprl}. The result reads
\begin{equation}\label{s3}
\Phi_\d(\sigma^z_3)=\sigma^y_{1,2} \sigma^y_{2,2} \sigma^y_{1,3}
\sigma^y_{2,3} ,
\end{equation}
so that the dual symmetry reduces to
\begin{equation}
\Phi_\d(\sigma^z_1\sigma^z_2\sigma^z_3)= \sigma^z_{1,2} \sigma^z_{2,2}
\sigma^z_{1,3} \sigma^z_{2,3}.
\end{equation}

The calculation of the last paragraph allows us to understand the
restrictions on the transverse-field bonds \(\sigma^z\) allowed in
\(H_{\sf HETC}\). That is,  the second term in Eq. (\ref{ely}) included
a sum over only one triangular sublattice.  The introduction of omitted
bonds such as \(\sigma^z_3\) on the other sublattice  would destroy the
exact dimensional reduction.  To see this, notice that for a model with
\(L_1,L_2\) arbitrary, the duality maps $\sigma^z_{L_1\a+\aa+\bm{b}}$
to 
\begin{eqnarray}
\Phi_\d(\sigma^z_{L_1\a+\aa+\bm{b}})= \prod_{i=2}^{L_1}\sigma^y_{1,i}
\sigma^y_{2,i}.
\end{eqnarray}
(Equation \eqref{s3} is a special case of this result, and similar
expressions hold for \(\Phi_\d(\sigma^z_{L_1\a+s\aa+\bm{b}}),\
s=2,\cdots,L_2\)). So, not only  the presence of
\(\sigma^z_{L_1\a+\aa+\bm{b}}\) in \(H_{\sf HETC}\) would introduce
long-range interactions in \(H_{\sf HETC}^D\), {\it but the range of
these interactions would grow without bound as a function of the area
on which \(H_{\sf HETC}\)} operates. This would spoil the exact dimensional
reduction in the thermodynamic limit. In contrast, adding
these bonds does not affect the {\bf d}$=1$-GLSs of the model. This
shows that {\bf d}-GLSs {\it can exist even in the absence of exact
dimensional reduction},  as already anticipated  in the introduction.

The same crucial feature can be seen from a different perspective. It is
intuitively clear that for {\it any} \(D\)-dimensional model one can
always devise artificially a dual, lower-dimensional, model. What sets
most of these dualities apart from true exact dimensional reduction is
again that {\it the range} of the interactions in the lower-dimensional
model will in general {\it grow without bound} as a function of the size
of the higher-dimensional model.   Only dualities that fail to show this
feature establish exact dimensional reduction.

Thus, we conclude this section by highlighting its important
result: the mapping of the $D=2$ quantum system of Eq. (\ref{ely})  on a
honeycomb lattice,  with a staggered field appearing only on one
triangular sublattice,   into the $d=1={\bf d}$ (ladder) short-range
system of Eq. (\ref{dual_ely}).

\subsection{Topological color codes}
\label{qcd}

We next briefly remark on dimensional reduction in the case of {\em
color codes}   \cite{color}.  The discussion will be very similar to
that in previous section.

Color codes are defined on planar graphs such that three and only three
links meet at any site of the graph (trivalent graphs), and such that
any plaquette is bounded by an even number of links. There are two
stabilizers per plaquette $p$: $B_{p}^{\alpha} = \prod_{i \in p}
\sigma_{i}^{\alpha}$ with $\alpha = x$ or $z$. Thus, the operators
$B_{p}^{x,z}$ are products of spins $S=1/2$ at lattice sites that belong
to a given plaquette. The total Hamiltonian is given by a sum of these
operators (bonds) over all plaquettes
\begin{eqnarray}
H_{\sf TCC} = - \sum_{p} B^{x}_{p} - \sum_{p} B^{z}_{p}.
\label{colorc}
\end{eqnarray}
We will not repeat the calculations explicitly  for this model, they are
analogous to those carried out for the model investigated in  Section
\ref{etch} and the $D=3$ model to be investigated in the next section,
Section \ref{xyz_road}.  Since, by definition of the lattice supporting
a color code, all plaquettes share an even number of  common sites
\cite{color}, the bonds $B_{p}^{\alpha}$ and $B_{p'}^{\alpha'}$ commute
always,
\begin{eqnarray}
[B_{p}^{\alpha}, B_{p'}^{\alpha'}]=0,
\label{bbpp}
\end{eqnarray}
for all $p,p',\alpha,$ and $\alpha'$, and 
\begin{eqnarray}
 (B_{p}^{\alpha})^{2}= \mathds{1}.
 \label{bp2}
\end{eqnarray}
If the lattice is realized on the surface of a torus,  the bond algebra
is supplemented  by two additional  relations (aside from Eqs.
(\ref{bbpp}, \ref{bp2})). These are given by
\begin{eqnarray}
\prod_{p} B_{p}^{\alpha} =\mathds{1},
\label{pbp}
\end{eqnarray}
for $\alpha = x,$ and $z$.  That is, the product of all operators
$B_{p}^{x}$ over all plaquettes $p$ is unity (and the same holds true
for the  product of all stabilizer operators $B_{p}^{z}$).  The system
of Eqs. (\ref{colorc}, \ref{bbpp}, \ref{bp2}, \ref{pbp}) is defined on a
Hilbert space of dimension $2^{N_{s}}$ with $N_{s}$ the number of sites
of the system.  The problem, as formulated, has a bond algebra
isomorphic to that of two decoupled, periodic Ising chains in zero
transverse field (see Ref. \cite{NO} for an analogous duality for the
toric code model). Thus, this system has an identical spectrum to two
independent, periodic classical Ising chains.  

Color codes have {\bf d}\(=1\)-GLSs. In particular, the toric color code
just described commutes with any  string operator obtained by multiplying
all the spins \(\sigma^x\) or all the spins \(\sigma^z\) 
on the links of any closed curve \cite{color}.


\subsection{The {\rm XXYYZZ} model}
\label{xyz_road}

\begin{figure}[htb]
\centering
\includegraphics[angle=0, width=.38\columnwidth]{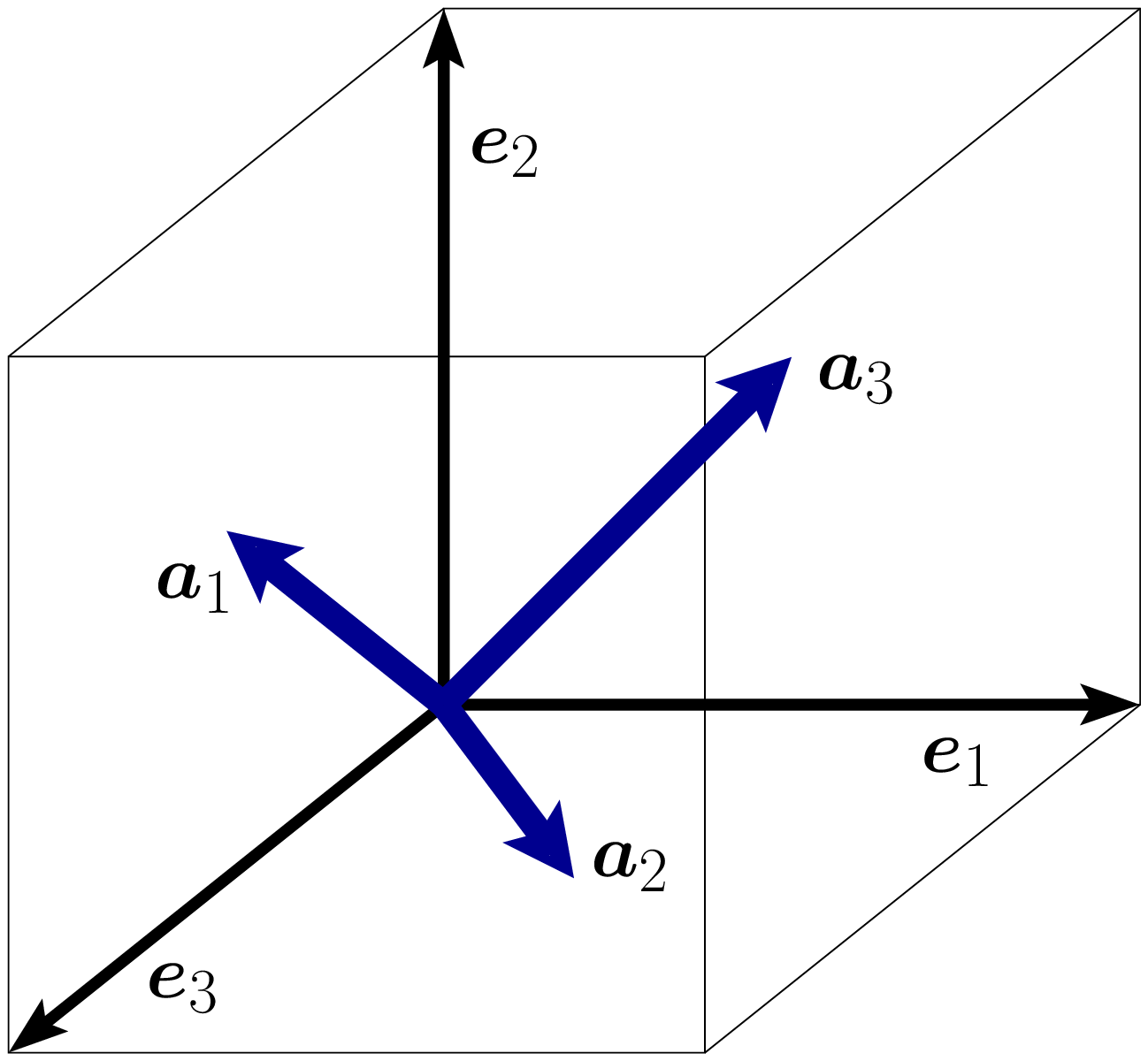}
\includegraphics[angle=0, width=.4\columnwidth]{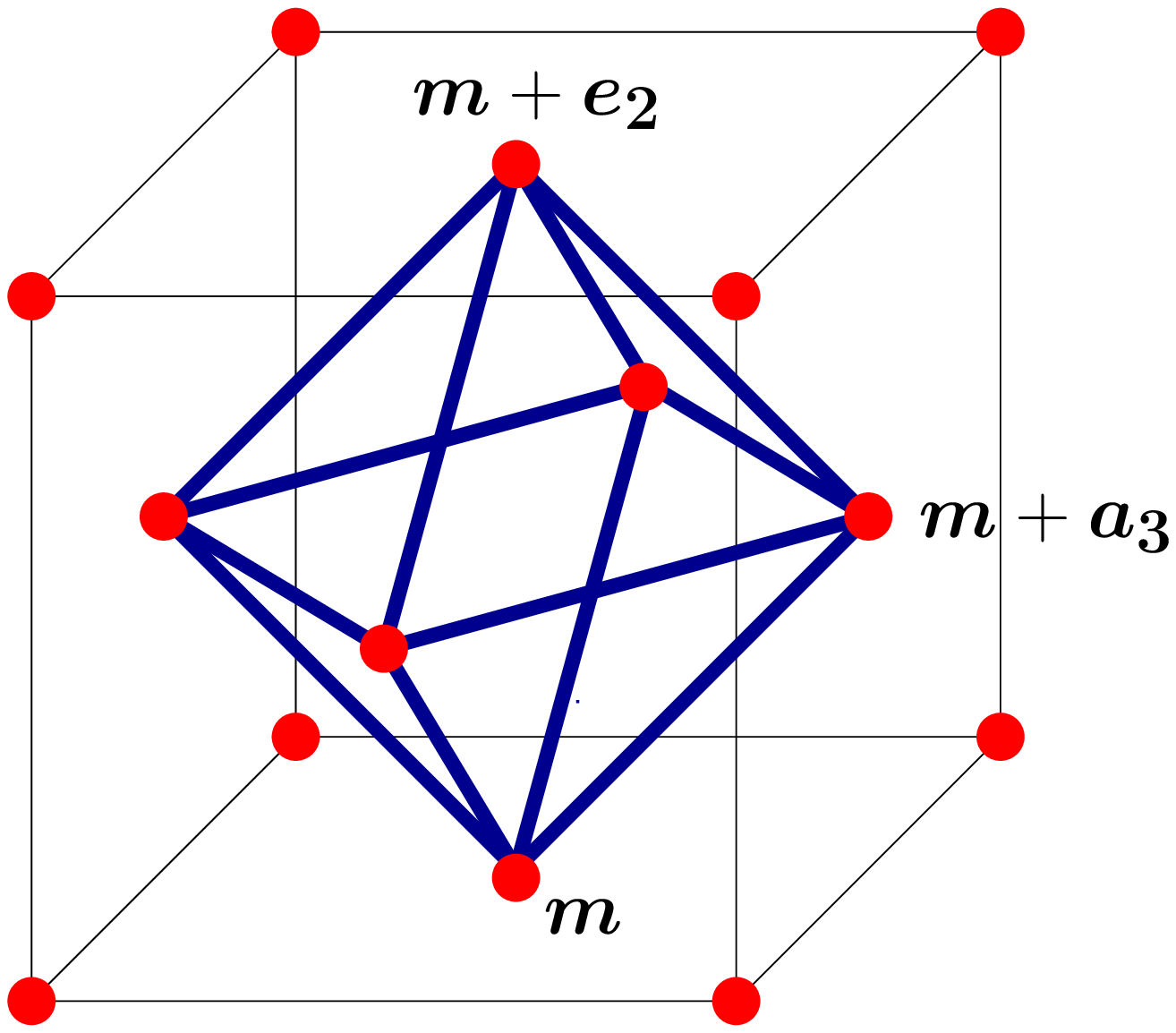}
\caption{The basis vectors \(\a,\aa,\aaa\) for the fcc Bravais lattice. 
For any site \(\m\) in an fcc lattice, there is one and only one
octahedron with \(\m\) as its lowest vertex.}
\label{a1a2a3}
\end{figure}

Our third example of dimensional reduction entails  a  system that we
call the  XXYYZZ model \cite{chamonmodel}. It can be described in terms
of  \(S=1/2\)  spins  placed on  sites  \(\m\) of an fcc Bravais lattice.
For concreteness, we can describe the sites  
\(\m=m^1\a+m^2\aa+m^3\aaa\), \(m^\mu  \in  \mathbb{Z}\) with
$\mu=1,2,3$,  as  linear combinations of the basis vectors, as illustrated in Fig. \ref{a1a2a3}, 
\begin{equation}\label{basis_vecs_fcc}
\a=\frac{\j+\k}{2},\ \ \aa=\frac{\k+\i}{2},\ \
\aaa=\frac{\i+\j}{2} .
\end{equation} 

The distinguishing feature of the fcc lattice that we would like to
exploit is the fact that it has exactly one elementary octahedron
per lattice site. In other words, for any site \(\m\) of the fcc
lattice, there is exactly one elementary octahedron having \(\m\) as,
say, its lowest vertex (and with every other vertex in the lattice as
well). Then, we can introduce the operator
\begin{equation}\label{def_O}
O_\m=\sigma^x_{\m+\a-\aa}\sigma^x_{\m+\aaa} 
\sigma^y_{\m}\sigma^y_{\m+\j}\sigma^z_{\m+\aaa-\aa}\sigma^z_{\m+\a},
\end{equation}
associated with each site in the lattice (notice that \(\i, \j, \k\) are
integral combinations of \(\a,\aa, \aaa\), and so they represent
displacements in the fcc lattice, even though they do not form a
basis).  In other words, the interaction terms are given by  the product
of all sites surrounding a given site of the even sublattice with the
product being of the form ``XXYYZZ''  wherein the component of the spin
appearing in the product is determined by its  location relative to the
center of the octahedron formed by the six sites.  This constitutes a
$D=3$ generalization of the ``XXYY" product appearing in  Wen's $D=2$
model   \cite{wenmodel} (that is also dual to Kitaev's Toric code
model   \cite{kitaev}).

\begin{figure}[htb]
\centering
\includegraphics[angle=0, width=.5\columnwidth]{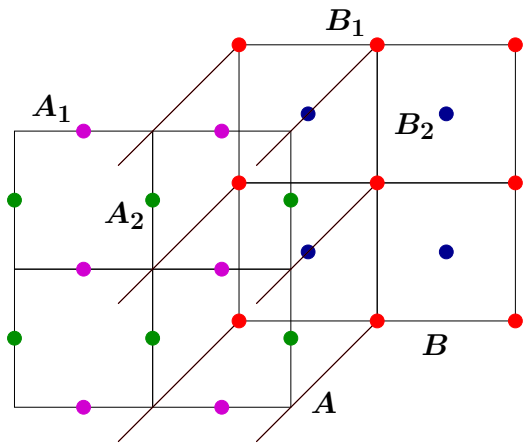}
\caption{The fcc Bravais lattice can be built as an $ABAB$ stack of two
square lattices ($A$ and $B$), displaced relative to each other (we show
a lattice elongated in the stacking direction to make the figure more
clear). Since the  planar lattices $A$ and $B$ are both bipartite, the
sites of an fcc lattice can be grouped into four disjoint classes. Any
two octahedra belonging to sites in the same class share at most one
vertex.}
\label{2_planes}
\end{figure}

Next, we introduce periodic boundary conditions in all three directions
\(\a,\aa,\aaa\), to obtain a finite fcc lattice \(\Lambda^P_{\sf fcc}\)
with  the topology of a  three-dimensional torus. More specifically, we
{\it identify} any two points \(\m\) and \(\m'\) equal up to a multiple 
of the period \(L_\mu\) in the \(\bm{a_\mu}\) direction. The fcc
structure forces the periods \(L_\mu\) to be {\it even} integers as
should become  clear below (see Fig. \ref{2_planes}). The Hamiltonian of
the XXYYZZ model reads
\begin{equation}
H_{\sf XXYYZZ}=-J\sum_{\m\in \Lambda^P_{\sf fcc}}\ O_\m.
\label{xyzxyz}
\end{equation}
This Hamiltonian acts on a state space of even dimension \(2^{N_s}\),
with \(N_s= L_1L_2L_3\) an integer {\it multiple of four}, and displays
{\bf d}\(=1\)-GLSs that are described in Ref. \cite{chamonmodel}. Our
goal is to employ the general techniques introduced in   \cite{bondprl}
to look for an exact duality of \(H_{\sf XXYYZZ}\) {\it to a
lower-dimensional model}.

The XXYYZZ model shares several features of typical models displaying
TQO: its bonds \(O_\m\) commute with each other and the degeneracy of
the ground  energy level depends on the topology of the underlying
lattice (that we  have fixed to be toroidal here). The bonds \(\{O_\m\
|\ \m \in \Lambda^P_{\sf fcc}\}\) generate a commutative bond algebra.
We saw already that dualities are encoded in the structure of this bond
algebra, so we need to specify it completely. The extra information we
need  is that, by virtue of the periodic boundary conditions, the bonds
\(O_\m\) satisfy four constraints. 

To understand this,  notice that we can split the sites of an fcc
lattice into four disjoint classes (at least for periodic and open
boundary conditions).  The reason is that, as shown in Fig.
\ref{2_planes}, the fcc lattice can be constructed as a stack \(ABAB\)
of  bipartite square lattices \(A,\, B\) displaced relative to each
other. Thus we can distinguish four types of  sites, according to
whether they belong to the sublattice \(A_1\) or \(A_2\) of an \(A\)
plane, or  \(B_1\) or \(B_2\) of a $B$ plane (see Fig. \ref{2_planes}). 
On the other hand, the bonds \(O_{\m_{\kappa}}\) in any one class
\(\kappa=A_1,A_2,B_1\), or  \(B_2\) share exactly one site with its
neighbors in the same class (and the corresponding spin, see Eq.
\eqref{def_O}). It follows that
\begin{equation}
\prod_{\m_{\kappa}}\ O_{\m_{\kappa}} =\mathds{1}.
\end{equation}
These are the four constraints that complete the specifications of  the
commutative bond algebra for \(H_{\sf XXYYZZ}\).

As explained in Sec. \ref{etch}, a Hamiltonian  will be dual to \(H_{\sf
XXYYZZ}\) if the two models share bond algebras with identical
structures (isomorphic bond algebras), and act on state spaces of the
same dimension. Let us consider then the model 
\begin{equation}
H_{\sf 4IP}=-J\sum_{\kappa=1}^4\sum_{m=1}^{N_s/4}
\sigma^z_{\kappa,m}\sigma^z_{\kappa ,m+1},
\label{4ip}
\end{equation}
that represents four independent (decoupled), periodic Ising chains. 
The bond algebra generated by the set of bonds 
\(\{\sigma^z_{\kappa,m}\sigma^z_{\kappa,m+1}\ |\ \kappa=1,2,3,4;\
m=1,\cdots, N_s/4\}\) is characterized by the fact that all bonds
commute, and satisfy four independent constraints
\begin{equation}
\prod_{m=1}^{N_s/4}\
\sigma^z_{\kappa,m}\sigma^z_{\kappa,m+1}=\mathds{1},\ \ \ \  \forall
\kappa.
\end{equation}
Also, \(H_{\sf 4IP}\) acts on a state space of dimension
\((2^{N_s/4})^4=2^{N_s}\). It follows that \(H_{\sf 4IP}\) is dual
(unitarily equivalent) to \(H_{\sf XXYYZZ}\). To establish a duality
mapping, let \(\kappa=A_1,A_2,B_1,B_2\) correspond to 
\(\kappa=1,2,3,4\), and let \(\phi_\kappa\) be any numbering of the
sites \(\m_\kappa\) in the sublattice \(\kappa=A_1,A_2,B_1,B_2\), from
one to \(N_s/4\).  Then
\begin{equation}
\label{omm}
O_{\m_\kappa}\dual \sigma^z_{\kappa,\phi_\kappa(\m_\kappa)}
\sigma^z_{\kappa,\phi_\kappa(\m_\kappa)+1}
\end{equation}
is a duality isomorphism that maps $H_{\sf XXYYZZ}$ to $H_{\sf 4IP}$, 
i.e., \(\Phi_\d(H_{\sf XXYYZZ})=\mathcal{U}_\d H_{\sf
XXYYZZ}\mathcal{U}_\d^\dagger=H_{\sf 4IP}\). {\it It follows that the
{\rm XXYYZZ} model must display exact one-dimensional behavior}. Thus,
in conclusion, we mapped the $D=3$ system of Eqs. (\ref{def_O},
\ref{xyzxyz}) on an fcc lattice to the $d=1={\bf d}$ system of Eq.
(\ref{4ip}) representing four decoupled Ising chains.

It is important to stress that the source of the dimensional reduction
in these two last models is {\it not} the completely commutative
structure of their bond algebras.  For instance,  a model as simple as the $D=2$ Ising model in
zero magnetic field
\begin{eqnarray}
H_{\sf I}=-\sum_{\mu=1,2}\sum_\r\ J_\mu \,
\sigma^z_\r\sigma^z_{\r+\bm{e_\mu}},
\end{eqnarray} 
with \(\r=r^1\i+r^2\j\), $r^\mu \in \mathbb{Z}$,  denoting the sites of
a simple square lattice, displays a completely commutative bond algebra
{\it but no dimensional reduction}, due to the presence of constraints
\begin{eqnarray}
&&\mathds{1}= \sigma^z_\r\sigma^z_{\r+\i}\times 
\sigma^z_{\r+\i}\sigma^z_{\r+\i+\j} \times
\sigma^z_{\r+\i+\j}\sigma^z_{\r+\j}\times \sigma^z_{\r+\j}\sigma^z_\r.
\nonumber
\end{eqnarray}
Notice also that the Ising model has no {\bf d}-GLSs with ${\bf d} < 2$.

\subsection{Effective boundary theories}

In the exact dimensional reduction mappings that we studied  thus far in
this section, the dimension of the Hilbert space was preserved. Thus,
for instance, when mapping a two-dimensional system onto a
one-dimensional one, the number of lattice sites was unaltered. By
contrast, in Section \ref{Q}, we illustrated how local boundary theories
can be constructed wherein the system size was reduced; these local
boundary theories, however, only provided bounds on the correlation
functions and, generally, did not constitute exact duals to the original
high-dimensional theory. In this brief subsection, we wish to briefly
point out that in  some cases local boundary theories can constitute
{\em exact} duals to the bulk system.  In such instances, the general
approach of exact dualities that we considered earlier in Section
\ref{exact_reduction} coincides with  that of the boundary theories of
Section \ref{Q}.  

Towards that end, we seek an effective Hamiltonian $H_{\sf eff}$ such
that 
\begin{eqnarray}
{\cal Z} = \tr_{\Lambda} (\exp(-\beta H))= {\cal N} \ \tr_{\Gamma}
(\exp(-\beta H_{\sf eff})).
\end{eqnarray}
Here, $H$ denotes the full Hamiltonian of the $D$-dimensional system
$\Lambda=\Gamma \cup \bar{\Lambda}$, whereas $H_{\sf eff}$ denotes an
effective Hamiltonian on a $d$-dimensional  boundary $\Gamma$.  An
overall normalization  factor is indicated by ${\cal N}$. 

In the quantum spin lattice models that we considered above, the trace
on any set of chosen bulk fields (or  {\it bonds} appearing in the
Hamiltonian   \cite{bond, bondprl}) can be done trivially. In the case
of the system of Eq. (\ref{ely}) in the absence of an external field
($h=0$) as well as the systems of Sections \ref{xyz_road}, and
\ref{qcd}, a partial trace over some of the bonds renders the  remaining
system to be that of decoupled bonds (or Ising spins in this case). In
particular, the effective boundary theories that contain all bonds that 
have their support on $\Gamma$ alone is given by that of decoupled Ising
spins,
\begin{eqnarray}
H_{\sf eff} = - \sum_{\x \in \Gamma} \sigma^{z}_{\x}.
\label{heffs}
\end{eqnarray} 
That is, the trace over the bulk fields $\bar{\Lambda}$ gives rise to a
Hilbert space volume factor alone that multiplies the partition function
associated with $H_{\sf eff}$ of Eq. (\ref{heffs}).  Now, the theory of
Eq. (\ref{heffs}) were to result from the  system of Eq. (\ref{ely}) in
the absence of an external field ($h=0$) and the systems of Sections
\ref{xyz_road} and \ref{qcd}, if we were to ``freeze'' out the bulk
spins in a spirit similar to that the derived local bounding  boundary
theories of Section \ref{Q}.

\section{Consequences of dimensional reduction: limits on the storage of
quantum information}

In general, we may apply the bounds of   \cite{NO,BN} to arrive at both
upper and lower bounds on all correlation lengths and autocorrelation
times in general systems with $\bf d$-GLSs. In particular, these bounds 
on autocorrelation times provide limitations on quantum memory schemes
and illustrate that, in some cases such as that of Kitaev's toric code
system \cite{kitaev},  information will be lost after a period that does
not scale with the system size. 
 
Kitaev's toric code model is a $D=2$ system defined on a torus (as the
name suggests) that exhibits a ${\bf d}=1$ $\mathbb{Z}_2\times
\mathbb{Z}_2$ symmetry.  The degrees of freedom in this model are spin
$S=1/2$ operators $(\sigma_{ij}^{x}$ or $\sigma_{ij}^{z}$) on the links
$(ij)$ of a square lattice having $|\Lambda|=N_{s}$ vertices that is
endowed with periodic boundary conditions. (To conform with the standard
conventions, we denote the links by $(ij)$ rather  than the particular
dual lattice site on which they are centered.) Specifically, this model
\cite{kitaev} is defined by the Hamiltonian 
\begin{eqnarray}
\label{HK}
H_{\sf Kitaev} = -\sum_{s=1}^{N_{s}} A_{s} - \sum_{p=1}^{N_{s}} B_{p},
\end{eqnarray}
with the {\it star} and plaquette operators $A_{s}$ and $B_{p}$ given by
\begin{eqnarray}
\label{AB}
A_{s} = \prod_j  \sigma_{js}^{x}, ~\  B_{p} = \prod_{(ij) \in p}
\sigma_{ij}^{z}.
\end{eqnarray} 
In the first of Eqs. (\ref{AB}), $j$ denotes all nearest neighbors of
$s$.  In Kitaev's model, information is stored in two ``{\it topological
qubits}'' determined by the eigenvectors of the ${\bf d}=1$  symmetry operators
\begin{eqnarray}
\label{qubit}
Z_{1,2} = \prod_{(ij) \in C_{1,2}} \sigma^{z}_{ij} , ~~ X_{1,2} =
\prod_{(ij) \in C^{\prime}_{1,2}} \sigma^{x}_{ij},
\end{eqnarray}
where $C_{1} (C_{2}^{\prime})$ are horizontal and $C_{2}
(C_{1}^{\prime})$ vertical closed  (${\bf d} =1$) contours on the
lattice (dual lattice) on the torus.  As illustrated in  \cite{NO},
Kitaev's model is special in that it may be exactly  mapped to two
decoupled classical Ising chains (i.e., a system of dimension  $d={\bf
d}=1$). Such a mapping is performed along lines similar to those of
Section \ref{exact_reduction}. The operators $A_{s}$  and $B_{p}$ map
onto bonds of the two decoupled Ising chains (each of length $N_{s}$).
The mapping is achieved by replacing, in the sum of Eq. (\ref{HK}),
$A_{s} \to \sigma^{z}_{s} \sigma^{z}_{s+1}$  on one chain and $B_{p} \to
\tau^{z}_{p} \tau^{z}_{p+1}$ on the other chain. That is, the
Hamiltonian of the $D=2$ dimensional system of Eq. (\ref{HK}) is dual  to that of
two Ising chains ($d={\bf d}=1$)
\begin{eqnarray}
\label{2IC}
H_{{\sf{2}\sf{Ising-chain}}} = - \sum_{s=1}^{N_{s}} \sigma^{z}_{s}
\sigma^{z}_{s+1} - \sum_{p=1}^{N_{s}} \tau^{z}_{p} \tau^{z}_{p+1},
\end{eqnarray}
with $\sigma^{z}_{N+1}= \sigma^{z}_{1}$ and $\tau^{z}_{N+1} =
\tau^{z}_{1}$. Similarly, within this mapping, the topological qubits
$Z_{1}$ and $Z_{2}$ map onto  single spins (say those located at the
sites number 1 on both chains), $Z_{1} \to \sigma^{z}_{1}$ and $Z_{2}
\to \tau^{z}_{1}$. The above set of mappings preserves all of the
algebraic relations amongst the {\it bonds} $\{A_{s}\}$ and $\{B_{p}\}$ 
that appear in the Hamiltonian of Eq. (\ref{HK}) as well as the two
qubits $Z_{1}$ and $Z_{2}$.

We now discuss consequences of dimensional reduction on the storage of
quantum information. First, we very briefly discuss how, physically, 
information may be lost by defects in this particular example system of
Kitaev's model.  Then, we turn to a formal framework that illustrates
how bounds on the storage of quantum information follow from our 
EQDR theorem. Lastly, we remark how such dimensional reductions allow
for an understanding of particular cross-over temperatures in finite
size systems. 

Conceptually understanding how dimensional reduction, when it occurs,
may limit the storage of quantum information is relatively
straightforward when thought about  in terms of defects that eradicate
the character  of an initial stored state and lead to a finite
correlation length. In Kitaev's toric code model, dimensional reduction
implies that topological defects that  arise in one-dimensional Ising
systems rear their head anew on the $D=2$ torus.  Similar considerations
apply for the other systems discussed in Section  \ref{exact_reduction}.
In an Ising chain the proliferation of domain walls leads, at any
positive temperature,  to a finite correlation length and to finite
autocorrelation (or memory) times.  In accord with the one-dimensional
character, the  low temperature autocorrelation times may  scale as
$\tau \sim \exp[\beta \Delta]$ where $\Delta$ is an energy penalty
associated with a domain wall.  In a one-dimensional Ising chain  with
exchange constant $J$, this penalty is system size ($N_{s}$) 
independent and set solely by the energy scale $J$. (In Eq. (\ref{2IC}),
the nearest-neighbor  exchange constant $J=1$.) Thus, any information
that is initially stored in Kitaev's toric code model  will, at
sufficiently long times, be destroyed by thermal fluctuations
\cite{NO}. 

We now step back to consider the more general consequences of effective
dimensional reductions borne by symmetries (Section \ref{Q}). (In
Kitaev's toric code model that exhibits the ${\bf d}=1$ GLSs of Eqs.
(\ref{qubit}) as well as  other systems in Section
\ref{exact_reduction}, these effective dimensional  reductions become
exact.) To obtain bounds on memory times,  we set the operator function 
$f$ of Section \ref{Q} to represent an autocorrelation function. That
is, we consider what occurs when we choose $f= Z_{1}(0) Z_{1}(t)$ so
that its expectation value will measure the autocorrelation of an
initial quantity $Z_{1}$ with itself at later times $t$. If quantum
memories are stable then $\langle f \rangle^D$ does not decay at long
times. By the results of Section \ref{Q}, the autocorrelation  function
in a general system with ${\bf d}$-GLSs, is bounded (both from above and
from below) by autocorrelation functions of a $d$-dimensional system in
which the range of the interactions and symmetries are preserved
\begin{eqnarray}
G_{\sf lower}^{(2)} \le G^{(2)}(t) \equiv \langle Z_{1}(0) Z_{1}(t) 
\rangle \le G_{\sf upper}^{(2)} ,
\end{eqnarray}
with $G_{\sf lower}^{(2)}, G_{\sf upper}^{(2)}$ autocorrelation
functions  in a  $d$-dimensional system. These lower-dimensional
autocorrelation functions serve as upper and lower bounds on the
autocorrelation function in $D$ spatial dimensions. It is notable that
in systems with interactions of finite range and strength the
autocorrelation times of quantities not invariant under  ${\bf d}=1$
GLSs do not scale with the system size. Thus, in such systems, the
autocorrelation times are system size independent. In a similar
fashion,  system size dependent  autocorrelation times appear in the
ordered phase of low-dimensional  systems with long-range interactions
exhibiting a phase transition, such as ferromagnetic spin chains with
$1/r^{\lambda}$ interactions ($\lambda \le 2$)   \cite{AYH}. Taken
together with the bounds of Section \ref{Q}, this allows for divergent
autocorrelation times of quantities that are not invariant under ${\bf
d}$-GLS in systems having long-range interactions. It is notable that
even though excitations may involve a macroscopic number of  spatially
local degrees of freedom in systems such as Kitaev and  the XXYYZZ
model, and thus it might be expected that the system exhibits glass-like
characteristics \cite{chamonmodel}, what matters in the computation of
the autocorrelation functions is not how memory may be erased in real
space. Rather, what is of importance,  so long as the noise is random,
is the algebra underlying these excitations. Thus, we would like to
suggest that, at least  for some realizations of noise stemming from an
external heat  bath, the XXYYZZ model exhibits one-dimensional memory
type effects and not  glass-like features.   

A peculiar feature of Kitaev's toric code model is the existence  of a
cross-over temperature which scales with the system size \cite{NO,CC}.
In what follows, we discuss this observation through the prism  of the
exact dimensional reduction to Ising chains that Kitaev's model
displays.  In an Ising chain with nearest-neighbor interactions (of
strength $J=1$) at an inverse temperature $\beta$,  the correlation
length scales as $\xi \sim - 1/\ln(\tanh \beta)$. This scaling implies
the existence of a crossover temperature scale $T_{\sf cross}$  below
which the entire chain is correlated and above which  the chain length
is larger than the correlation length. On a single Ising chain with
$N_{s} \gg 1$ spins, this gives rise to a crossover temperature $T_{\sf
cross} \sim 1/\ln N_{s}$   \cite{NO}. A similar crossover temperature is
seen when the entanglement entropy  is analyzed \cite{CC}.  A
rudimentary understanding of this scaling can be seen from standard
energy-entropy balance considerations in the free energy, $F = E - T{\sf
S}$, for the insertion of a ${\bf d}=1$ defect (i.e., a domain wall in a
one-dimensional  chain for discrete ${\bf d}= 1$ symmetries) in a system
with short-range interactions.   Associated with the insertion of  a
defect there is  an energy penalty  $E =\Delta$ and an entropy  ${\sf S}
\sim \ln N_{s}$. These two contributions lead to a a crossover
temperature scaling as $T_{\sf cross} \sim 1/\ln N_{s}$. For $T<T_{\sf
cross}$ defects are unfavorable, while $T>T_{\sf cross}$ defects
proliferate. The discussion above can replicated {\it mutatis mutandis}
for the  systems of Section \ref{exact_reduction} when these reduce to
Ising chains as well as many others which reduce to other 
low-dimensional systems. The existence of predicted crossover
temperatures in these  systems, as adduced from our general
considerations above regarding dimensional reductions,  may hopefully be
bolstered  by explicit computations of thermodynamic quantities
\cite{NO} or entanglement entropy  \cite{CC} in future works. 

\section{Conclusions}
\label{conc}

In conclusion, we examined both {(i)} {\it effective} and {(ii)}  {\it
exact} dimensional reductions and holographic correspondences. 
Specifically, we proved that 

{(i)} In a rather general local $D$-dimensional quantum theory that is
defined on some volume  $\Lambda$,  correlation functions involving
fields that lie on a subspace $\Gamma \subset \Lambda$ of dimension
$d<D$ can generally be {\em bounded} by correlation functions in a {\it
local} $d$-dimensional theory. These bounds lead to an {\it effective
dimensional reduction} which become most potent  when the original
$D$-dimensional theory exhibits a particular symmetry (or set of
symmetries)  having its support on the region $\Gamma$. (A symmetry
operator that has its support on a general region of dimension ${\bf d}$
is termed a ${\bf d}$-dimensional gauge-like  symmetry.) When the
symmetry operator acts on the ${\bf d} = d$ spatial region  $\Gamma$,
the resulting bounds of Eq. (\ref{strong}) lead to a generalized quantum
Elitzur theorem.

{(ii)} Stronger than effective dimensional reductions are {\it exact
dimensional reductions} that relate systems in different dimensions. We
studied and illustrated exact dimensional reductions of two different
types: \newline
{(a)} Dualities centered on the large-$n$ limit of $O(n)$ vector
theories and  the high temperature (or weak coupling) limit.  The
dualities derived in this case relied on the dependence of the partition
function (or generating functional) on a local effective density (in the
latter case, the density of states for single modes in the
non-interacting limit which retains its character also in the presence
of interactions in the large-$n$ limit or that of high temperatures). We
briefly speculated that it may be possible to search for analogues of 
functional theory type treatments \cite{kohn}  in field theories
and interacting many-body systems that effectively render the theory
nearly non-interacting. With the aid of these functionals,  it may be
possible to more generally construct transformations that do not deform
the appropriate {\it functions} and thus preserve the features of a
local theory when it is related to another local system in a different
spatial dimension.  \\
{(b)} We further examined exact dualities that do not appear in some
limit but that, rather, always hold true. The studied dimensional 
reduction dualities focused solely on the physical {\it bonds} that
appear in the Hamiltonian or action defining the theories. With the aid
of this approach, redundant degrees of freedom on both sides of a
general duality (e.g., {\it gauge symmetries} in various gauge theories,
{\it local coordinate invariance} in gravity theories, etc.)  may be
automatically discarded.  The equivalences in case {(ii)} focus on the
form of the spectrum of the theory:  case {(a)} focuses  on the local
density of states (in, for instance, $k$-space) and {(b)} on mappings that 
trivially preserve the exact spectrum of the theory. 

Both the {(i)} effective and {(ii)} exact  dimensional reductions
introduced in this work {\it do not require an actual physical
compactification of dimensions}. 

\section{Acknowledgements}
We are indebted to inspiring discussions with Pawel O. Mazur, and Emil
Mottola at the {\it Condensed matter meets gravity} workshop in the
Lorentz Institute, Leiden, in August 2007 and to illuminating interactions with
Clifford M. Will and Emanuel Knill.  We wish to thank Simin Mahmoodifar for a very careful
reading of the manuscript and corrections.  ZN is indebted to the
National Science Foundation (DMR-1106293) for support. GO acknowledges
partial support by the National Science Foundation under  Grant No.
1066293 and the hospitality of the Aspen Center for Physics.

\appendix
\section{Entanglement-based version of the EQDR theorem}
\label{alternativeeqdr}

We briefly describe a variant of the EQDR theorem that holds for 
arbitrary states \(\rho\), at least if the Hilbert space
of the system is finite-dimensional. 
The assumptions are again that we
can treat the bulk and the boundary as distinguishable, 
\(\mathcal{H}_{\Lambda}=
\mathcal{H}_{\Gamma}\otimes\mathcal{H}_{\bar{\Lambda}}\), and 
that operators \(f\) localized on \(\Gamma\) have the form
\(f=f_\Gamma\otimes \mathds{1}_{\bar{\Lambda}}\). Then
we can expand {\it an arbitrary state operator} \(\rho\) as 
\begin{equation}\label{sumot}
\rho=\sum_{i} \lambda_i \, \rho_{\Gamma i}\otimes\rho_{\bar{\Lambda} i},
\end{equation}
where \(\rho_{\Gamma i}, \rho_{\bar{\Lambda} i}\) are 
state operators,  and \(\lambda_i\) {\it may be any real number, positive
or negative} such that \(\sum_i\lambda_i=1\). Separable or unentangled 
states correspond to the case where all $\lambda_i$ happen to be positive.
An entangled state is characterized by the presence of at least one
negative coefficient in  \eqref{sumot}.

Let us first justify Eq. \eqref{sumot}. 
Notice that an arbitrary Hermitian operator can be 
written as the difference of two positive operators, by virtue of the 
spectral decomposition theorem. If \(O\) is Hermitian and \(o_i, P_i\) are its 
real eigenvalues and
eigenprojectors, then
\begin{equation}\label{pos}
O=\sum_i o_iP_i=\sum_{i_+} o_{i_+}P_{i_+}-\sum_{i_-}|o_{i_-}|P_{i_-}=O^+-O^-,
\end{equation}
where the index
\(i_{+}\ (i_{-})\) varies over that subset of \(i\)'s for which 
\(o_i\) is positive (negative). We can identify
the real space of Hermitian linear transformations \(J_\Lambda\)  
on \(\mathcal{H}_{\Lambda}\) with the 
space \(J_\Gamma\otimes J_{\bar{\Lambda}}\), where \(J_\Gamma\) and 
\(J_{\bar{\Lambda}}\), represent transformations on the boundary and bulk, 
respectively. Since these are finite dimensional
spaces, it follows \cite{algebras} that any Hermitian operator on 
\(\mathcal{H}_{\Lambda}\), and any 
state operator \(\rho\) in particular, can be written as 
a real linear combination
\begin{equation}\label{gexp}
\rho=\sum_{i} O_{\Gamma i}\otimes O_{\bar{\Lambda} i}.
\end{equation}
While there is no guarantee that the \(O_{\Gamma i},O_{\bar{\Lambda} i}\) will
be state operators, we can apply the decomposition of Eq. \eqref{pos} to each
\(O_{\Gamma i},O_{\bar{\Lambda} i}\) to turn the expansion of Eq. \eqref{gexp}
into an expansion in terms of products of positive operators. If we next normalize
these positive operators, we get the expansion of Eq. \eqref{sumot}, and we 
see that the \(\lambda_i\), although real,  need not be positive by virtue of the minus sign
in Eq. \eqref{pos}. 

We can now use Eq. \eqref{sumot} to
compute inequalities of interest to dimensional reduction and holographies. 
If \(f=f_\Gamma\otimes \mathds{1}_{\bar{\Lambda}}\) is an observable 
\begin{eqnarray}
\label{entangledeqdr}
L_+\langle f\rangle^+_l-L_-\langle f\rangle^-_l\leq\tr_{\Lambda}(\rho f)
\leq L_+ \langle f\rangle^+_u -L_{-} \langle f\rangle^{-}_u.
\end{eqnarray}
where \(L_+=\sum_{i_+} \lambda_{i_+}\), \(L_{-}=\sum_{i_{-}}|\lambda_{i_{-}}|\)
are both positive, 
\begin{equation}
\langle f\rangle^+_u\equiv{\sf max}_{i_+}\tr_{\Gamma}(\rho_{\Gamma i_+}f_\Gamma),\ \ \ \ 
\langle f\rangle^{-}_u\equiv {\sf min}_{i_-}\tr_{\Gamma}(\rho_{\Gamma i_-}f_\Gamma),
\end{equation}
and 
\begin{equation}
\langle f\rangle^+_l\equiv{\sf min}_{i_+}\tr_{\Gamma}(\rho_{\Gamma i_+}f_\Gamma),\ \ \ \ 
\langle f\rangle^{-}_l\equiv {\sf max}_{i_-}\tr_{\Gamma}(\rho_{\Gamma i_-}f_\Gamma).
\end{equation}
Notice that if  the state \(\rho\) is unentangled, then \(L_-=0\) and $L_+=1$, 
and this version of the EQDR theorem, Eq. \eqref{entangledeqdr}, comes remarkably close 
to the classical dimensional reduction theorem, as perhaps is to be
expected. In general, the inequalities of Eq. \eqref{entangledeqdr}
clearly illustrate another way of achieving effective dimensional reduction.


\begin{thebibliography}{00}
\bibitem{ssobs}
In principle, the same superselection rules applies to observables.
However, when computing an average we have that
\begin{equation}
\tr_{\Lambda}(\rho f)=\tr_{\Lambda}(P_{\sf Fock}\rho P_{\sf Fock} f)=
\tr_{\Lambda}(P_{\sf Fock}\rho P_{\sf Fock} P_{\sf Fock}fP_{\sf Fock}),
\end{equation} 
and so even if an observable does not respect the superselection 
rule it will automatically be substituted for one that does, provided
the state \(\rho\) carries the correct quantum statistics.
\end{thebibliography}
\end{document}